\documentclass[%
 reprint,
 amsmath,amssymb,
 aps,
]{revtex4}

\usepackage{graphicx}% Include figure files
\usepackage{dcolumn}% Align table columns on decimal point
\usepackage{bm}% bold math

\usepackage{hhline}
\usepackage{setspace}
\usepackage{subfigure}

\begin{document}

\preprint{APS/123-QED}
\def\mean#1{\left< #1 \right>}

\title{Probing the losses for a high power beam}

\author{M.~Haj~Tahar} \email{malek.haj-tahar@psi.ch}%
\author{D.~Kiselev}
\author{A.~Knecht}
\author{D.~Laube}
\author{D.~Reggiani}
\author{J.~Snuverink}

\affiliation{
Paul Scherrer Institut, Villigen, Switzerland
}

\begin{abstract}
The High Intensity Proton Accelerator (HIPA) cyclotron at the Paul Scherrer Institut (PSI) delivers 590 MeV CW proton beam with a maximum power of 1.42 MW. After extraction, the beam is transferred in a 120 m long channel towards two target stations (TgM and TgE) for surface muon production before depositing its remaining power at the spallation target SINQ for neutron production. As part of the High Intensity Muon Beamline (HIMB) feasibility study, the first of these targets will be replaced with a thicker one thereby increasing the rate of surface muon production. However, a key challenge for HIMB is to maintain the proton beam losses to the lowest possible levels which requires improving our understanding of the distributed losses along the MW-class beamline. To this end, a new approach was developed where the aim is to relate the experimental values of the temperature, beam profile measurements as well as beam current measurements to the combined power deposition calculations and primary beam losses using Monte Carlo simulation tools.
\end{abstract}

\pacs{Valid PACS appear here}% PACS, the Physics and Astronomy
\maketitle

%\tableofcontents

\section{Introduction}

\begin{figure*}
\centering 
\includegraphics*[width=16cm]{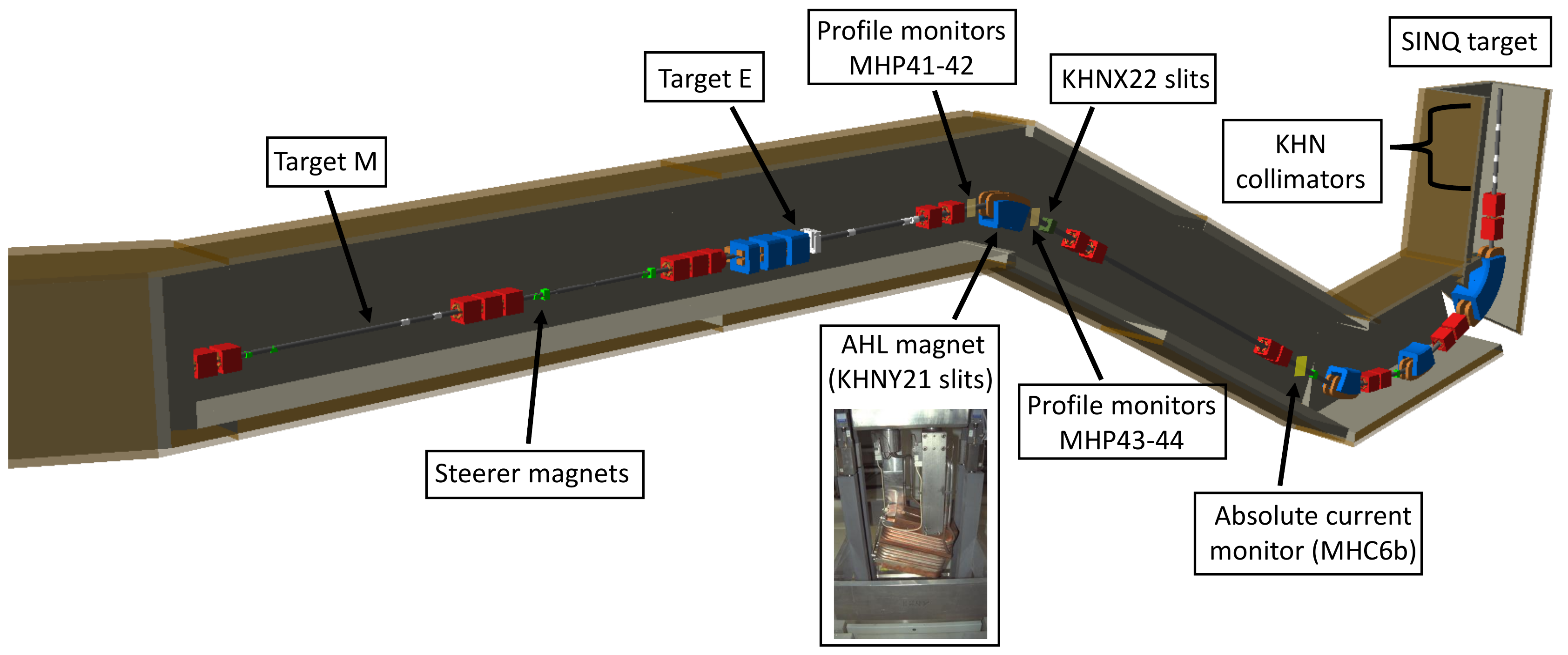}
\caption{BDSIM model of the beamline. The tunnel is implemented to guide the eye only while all shielding surrounding the beamline is disabled for clarity purposes. Note that the vertical beam halo scraper KHNY21 is located in the central region of the AHL magnet and that TgE collimators are visible in white just behind TgE. }
\label{fig:3D_model}
\end{figure*}

The High Intensity Proton Accelerator HIPA complex at PSI has been operating since 1974 with several major upgrades allowing to reach an average beam power of up to 1.42 MW thanks to a two-stage cyclotron \cite{grillenberger2021high}: the first stage takes place in injector II where the proton beam pre-accelerated by a Cockcroft-Walton DC linear accelerator to 0.87 MeV is accelerated up to 72 MeV, while the second stage taking place in the main ring cyclotron allows accelerating the beam up to 590~MeV with a maximum achieved average current of 2.4 mA.
After extraction from the main ring with an efficiency better than 99.98\%, the CW beam is transported in three different stages to impinge on three different targets \cite{kiselev2021meson} for meson and neutron production: The first stage of the proton beamline delivers the beam to a thin 5 mm target M (TgM): it consists of a 43 m long channel containing 5 bending magnets, 12 quadrupoles, 12 steerer magnets and 22 profile monitors. This section allows the transport of the high power beam with losses below the 1 W/m level. After impacting TgM, the beam is transported in a second stage to a thick 40 mm target~E (TgE). This section is 18 m long, consists of 2 quadrupole triplets, 3 bending magnets, 5 profile monitors as well as 2 horizontal and 2 vertical steering magnets. In the third and last stage, the beam is directed towards the Swiss Spallation Neutron Source (SINQ) target for neutron production: this section is 55 m long, consists of 7 collimators, 4 bending magnets, 12 quadrupoles, 9 profile monitors, 5 steering magnets and 3 pairs of slits. A schematic of the section between TgM and SINQ is displayed in Fig. \ref{fig:3D_model}.

To keep HIPA at the forefront of intensity frontier research, the IMPACT (Isotope and Muon Production using Advanced Cyclotron and Target technologies) project was recently proposed by Paul Scherrer Institut (PSI), the University of Zurich (UZH), and the University hospital of Zurich (USZ) with the objective to construct two new target stations and beamlines at PSI \cite{CDR_HIMB}. One of these target stations aims to produce a wide range of radionuclides in the required quantities for clinical studies related to the treatment of tumors \cite{TATTOOS}, while the second focuses on increasing the rate of surface muons by two orders of magnitude ($\sim 10^{10} \mu^{+}/s$) hence the name High Intensity Muon Beamline (HIMB) \cite{aiba2021science}. The goals for HIMB will be achieved through a series of measures: first, the existing target station M will be replaced with a newly designed target H (TgH) which will increase the effective thickness travelled by the proton beam in the graphite material from 5 mm to 20 mm. Second, the use of a slanted target geometry will allow a twofold increase in surface muon rates compared to the standard version \cite{berg2016target}. Last but not least, owing to the use of large-aperture solenoids and dipoles, the copious amounts of muons will be efficiently collected and transported, reaching an overall transmission of around 10\%. The study discussed in this paper, as well as all measurements, are in the frame of the HIMB feasibility study.

Despite the major benefit of increasing the thickness of TgM, a key challenge for HIMB is to maintain the primary proton beam losses to the lowest possible levels. This requires a comprehensive understanding of the distributed losses along the beamline. To this end, one of the major aims of this work is to benchmark BDSIM/GEANT4 \cite{nevay2020bdsim, agostinelli2003geant4} simulations of the high power SINQ beamline against some of its measured characteristics.
In addition, as a means to understand the losses for MW-class beams, we shall provide a profound discussion about the importance of combining the Monte Carlo simulation tools with an accurate model of the optics. \\
Although the physics models in Monte Carlo simulation tools are generally well benchmarked against other codes and/or experiments \cite{lo2015geant4, ALLISON2016186}, a validation against macroscopic quantities is deemed essential in order to assess the level of accuracy achieved from simulations \cite{lechner2019validation}. When the high power beam is brought in contact with the accelerator material, secondary particle showers will be created and power deposition will be induced. The latter is among the most important and often invoked quantities for high power beams since it is needed to perform the thermal analysis for the relevant components, design the shielding and devise an adequate scheme to protect the beamline (by means of appropriate collimation, cooling and machine protection system). Therefore, it is natural to probe the losses by means of temperature measurements.
Relying on such a macroscopic quantity, we shall devise an experimental approach to determine the beam deposited power and compare it with simulations. \\
Before embarking in our benchmarking campaign, the main considerations of the developed simulation model and the methodology to probe the losses will be discussed. A summary of the paper outline will follow then. 

\section{Main considerations: SIMULATION MODEL AND METHODOLOGY}

Simulating the scattering processes of the primary proton beam at the target is particularly challenging for high power accelerators where there is a strong need to limit the primary beam losses to the lowest achievable levels. Besides, the secondaries originating from the non-elastic nuclear interactions are crucial to accurately calculate the power deposition. For these reasons, we shall rely on Monte Carlo simulation tools. In addition, tracking particle distributions in a sequence of elements with various magnetic fields as commonly achieved in reference accelerator design codes such as MAD-X \cite{grote2003mad}, is crucial for our analysis. \\
Furthermore, one of the major aims of this paper is to understand and explain how the beam transmission to SINQ target evolves with the injected beam current: for this reason, a dedicated experimental campaign was performed in 2020-2021 in order to compare our simulations to the measurements. Given that the beam conditions from the cyclotron might change, all measurements (beam profiles, temperatures, beam current) were taken simultaneously. In what follows, we discuss the most important aspects of the simulation model.

\subsection{Monte Carlo simulation}

The above considerations lead to the choice of BDSIM as the reference program for all calculations: BDSIM combines the accurate accelerator tracking routines with the physics processes in the GEANT4 toolkit. The physics lists that are the basis of all our simulations are added together in a modular fashion and were selected based on the recommended ones from BDSIM/GEANT4: physicsList=``em ftfp$\_$bert decay muon hadronic$\_$elastic em$\_$extra''. \\  The hadron-nucleus interactions of interest (below 0.59 GeV incident proton energy) were described using the GEANT4 Bertini cascade model (ftfp$\_$bert) which is one of the most used reference physics lists for high energy applications~\cite{ALLISON2016186}. In addition, our GEANT4 model constructed by BDSIM relies on the recommended ``Matrix'' integration algorithm. For paraxial on-momentum particles a matrix-like step tracking is performed relying on the magnetic parameters, while for non-paraxial particles BDSIM resorts to a Runge-Kutta (G4ClassicalRK4) algorithm relying on 3D electromagnetic fields. This allows all particles to be tracked in all directions over all momentum ranges, but with the accuracy and speed of accelerator tracking for paraxial particles.

\subsubsection{Geometry and materials definition}

The choice of implementing the geometry as accurately as possible is justified by our aim to benchmark the power deposition calculations with the measurement along the SINQ beamline.
Implementing the shielding is equally important for the same purposes. Owing to the already built-in geometry models in BDSIM, a hybrid approach is privileged whereby the magnets are created from the pre-defined models in BDSIM, while the more complex elements, particularly those in the vicinity of TgE and SINQ, require a more dedicated effort. For instance, conical collimators with varying elliptical cross sections and segmented features were implemented by importing and converting the Computer-Aided Design (CAD) files to tesselated solids based on the Geometry Description Markup Language (GDML) \cite{GDML}. This benefited from the recently developed python library \textsc{Pyg4ometry} \cite{WALKER2022108228} which allows the implementation of quite complex geometries in a few seconds. The GDML format is particularly suitable for further editing of the geometry since it supports the addition of shielding volumes, which are in general not provided in the CAD files. In summary, our geometry was developed in a way to reproduce the specific details of the beamline in the region that suffers from the largest losses, while omitting some details that are irrelevant for our analysis, yet time-consuming, in the region with negligible losses. In doing so, care must be paid to ensure no overlaps exist since this can lead the particle to skip some geometries/volumes without any error generation. Scanning the geometry for errors is available to overcome this (checkoverlaps option), however, a more careful inspection of the power deposition calculation is often crucial to reveal the trailing errors that may not be detected otherwise. \\
Next, the material composition is assigned to each volume. This relies on the GEANT4 pre-defined materials from the NIST database. The main exception to this is the target material for which the density shall be carefully defined since it plays the most important role on the subsequent losses: TgM and TgE are made of polycristalline graphite. While the first target used at PSI in 1985 had a density of 1.7 g cm$^{-3}$, the most recent analysis performed in 2021 showed that the density of the target is 1.81 g cm$^{-3}$. This is still far from the value of 2.26 g cm$^{-3}$ for natural graphite, however, it shows the improvement of the manufacturing process over the years. \\
Last but not least, the beamline model is obtained by converting the TRANSPORT \cite{brown1980transport} input files into BDSIM gmad~\cite{agapov2006gmad} input ones and the 3D geometries are carefully placed as discussed earlier.  
A sketch of the beamline developed model between TgM and SINQ is displayed in Fig. \ref{fig:3D_model} where all shielding is removed for clarity purposes. 

\subsubsection{Optics}

\begin{figure*}
\centering 
\includegraphics*[width=12cm]{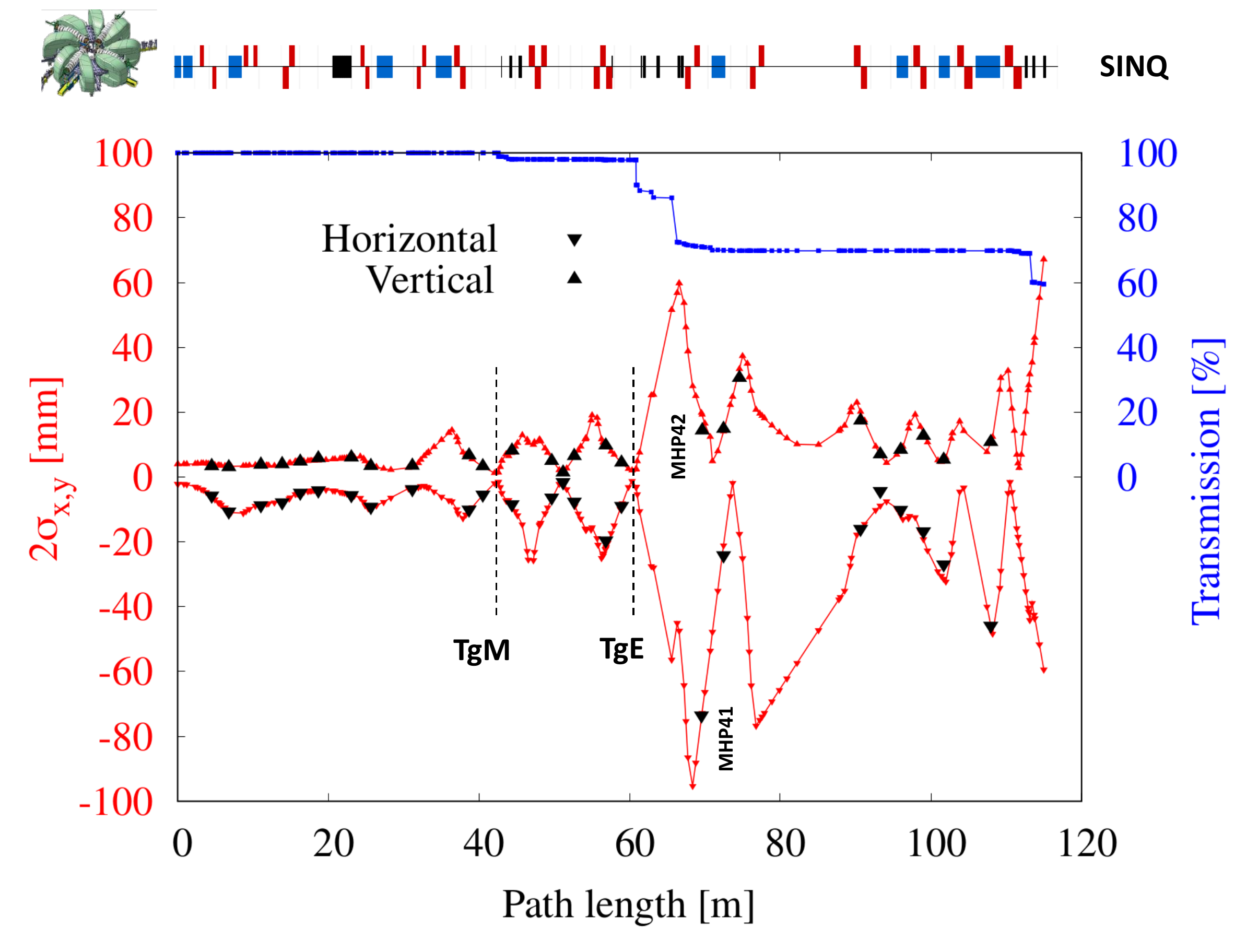}
\caption{Comparison of beam envelopes with the beam profile measurements. The beam transmission is also shown (right axis).}
\label{fig:optics}
\end{figure*}

\begin{figure}[!h] 
\subfigure{\includegraphics[width=6cm]{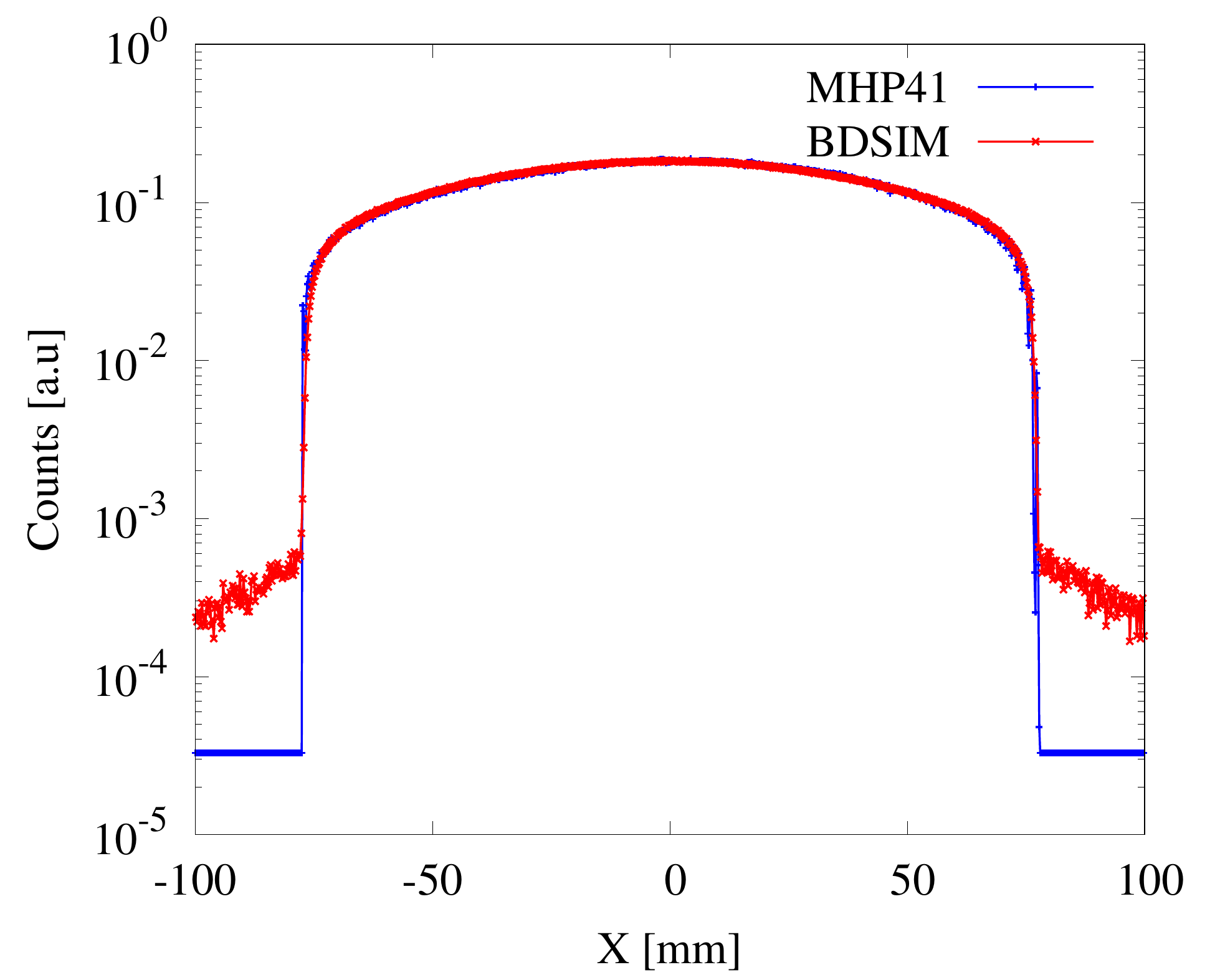}} \hspace{10mm} 
\subfigure{\includegraphics[width=6cm]{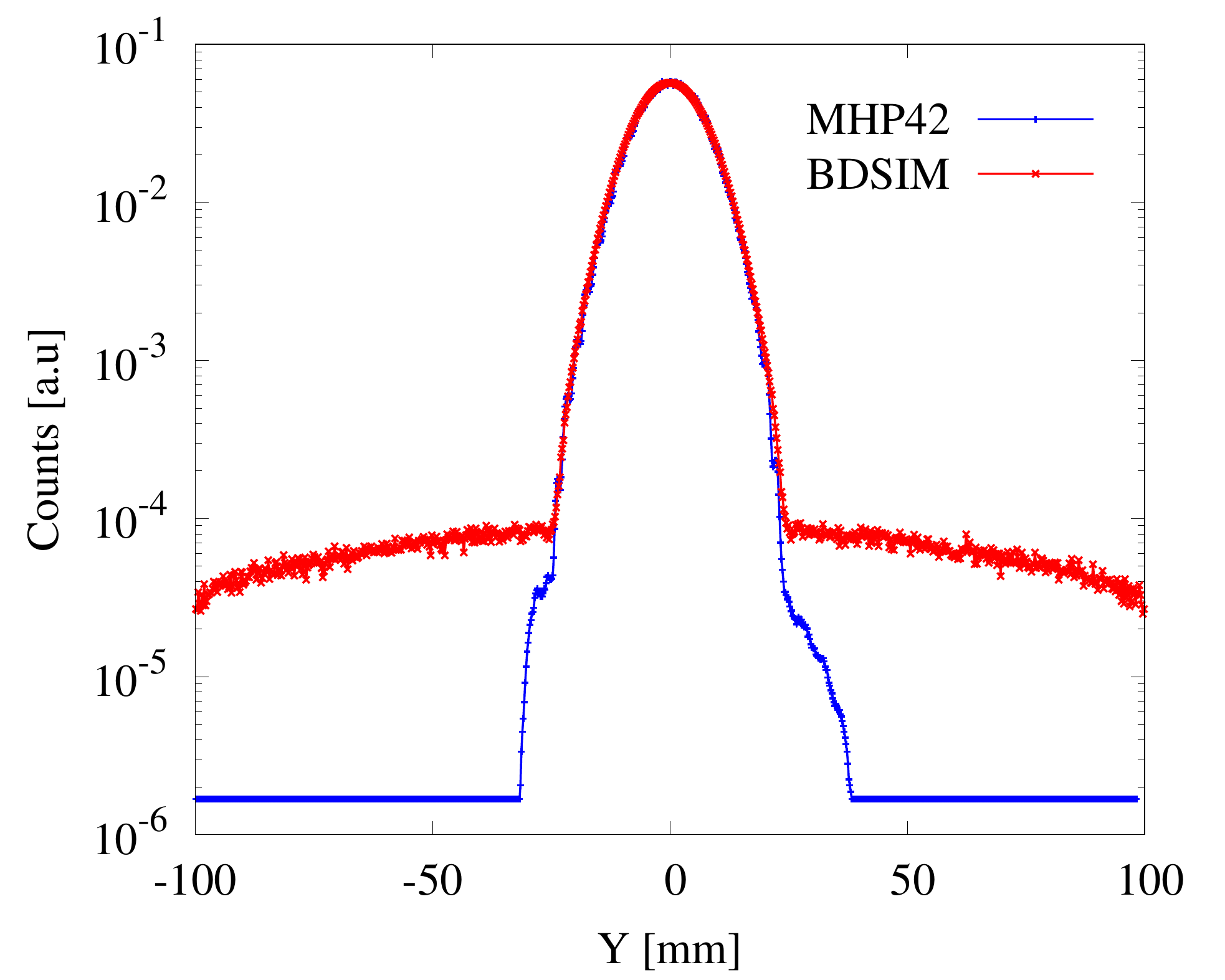}}
\vspace{5mm}
\subfigure{\includegraphics[width=6cm]{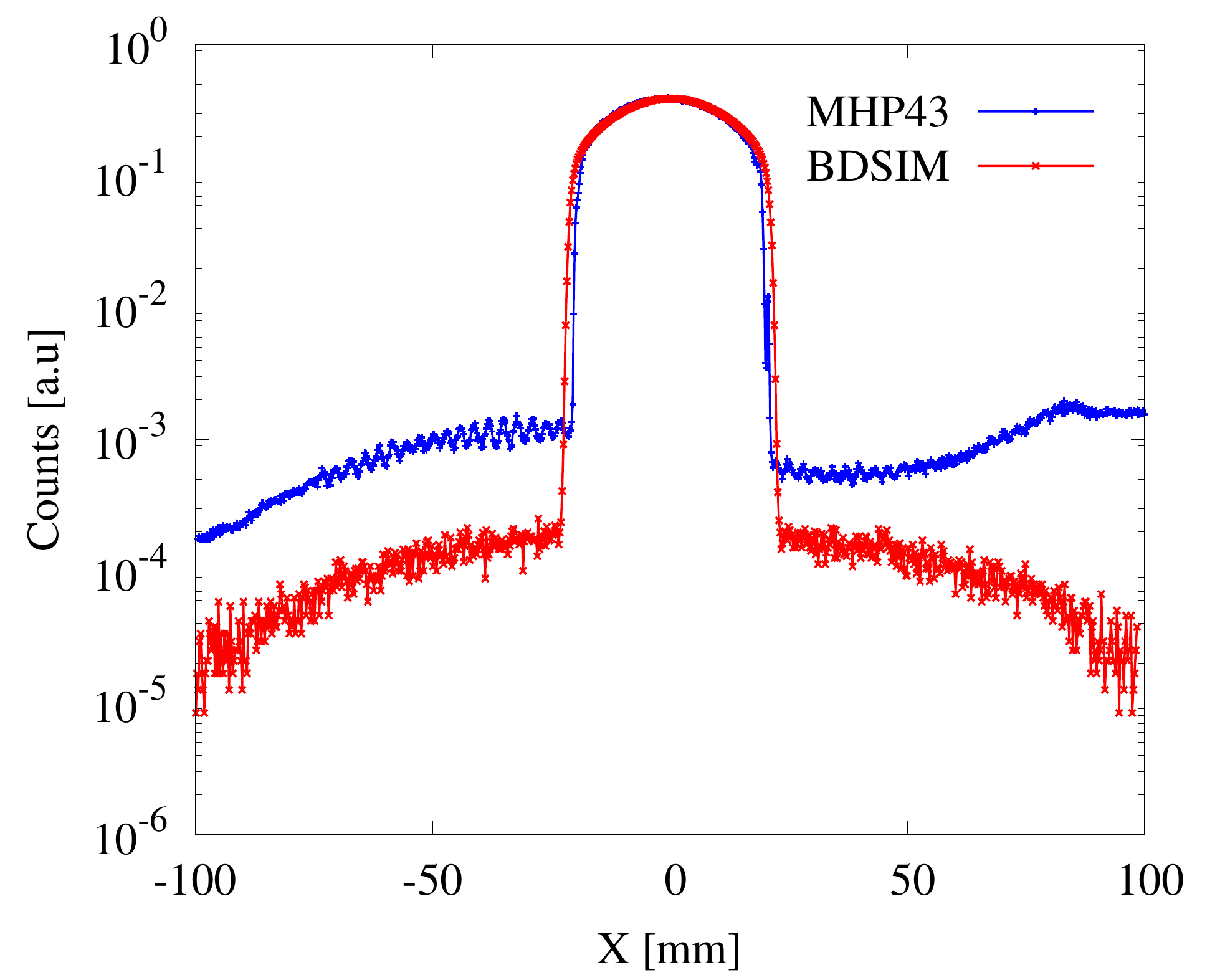}} \hspace{10mm}
\subfigure{\includegraphics[width=6cm]{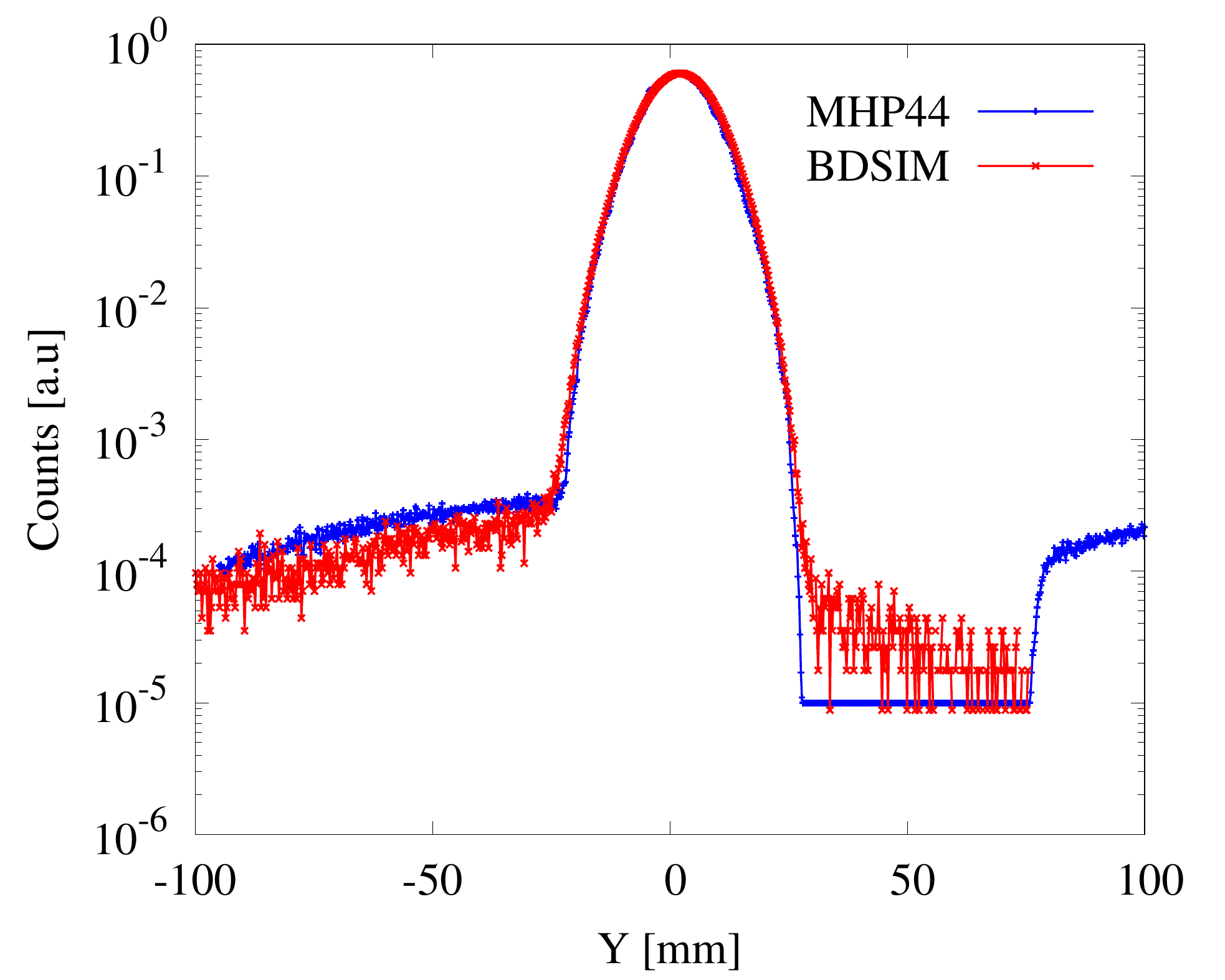}}
\hfill
\caption{Comparison of measured horizontal (left) and vertical (right) beam profiles with BDSIM simulations for a beam current of 1.9 mA. The measured profiles are shown in blue while the simulation results are displayed in red.}
\label{fig:MHP_comparison}
\end{figure}

Owing to its built-in fitting capability, TRANSPORT is used to determine the initial beam conditions at the extraction of the main ring cyclotron that matches the measured beam widths up to TgM. 
Such initial beam conditions are subsequently fed into BDSIM and tracking performed for the entire beamline from the ring extraction up to SINQ target. Comparison of the obtained beam envelopes with the measurements is shown in Fig. \ref{fig:optics} where a good agreement can be observed. The beam transmission drop with the current is also illustrated: in the first stage of the beamline (up to TgM), the losses are negligible. In the second stage, upon impacting TgM and cleaning its tails, 98~\% of the initial proton beam current reaches TgE. In the third and last stage, almost all losses aside from the KHN collimators occur between TgE and the KHNX22 slits. For this reason, it is particularly instructive to compare the measured beam profiles with the simulated ones in such a loss-dominated region. The comparison is shown in Fig.~\ref{fig:MHP_comparison} for MHP41-42-43-44 which can be spotted in Fig. \ref{fig:3D_model}. Several remarks are in order. First, the beam profiles are all but Gaussian in shape: after impacting TgE, the core of the angular beam distribution evolves into a Gaussian shape due to the multiple coulomb scattering, while the single scattering events lead to the formation of lateral tails \cite{rizzoglio2017evolution}. By means of TgE collimators, the beam is subsequently cut in the horizontal plane mainly to protect the beamline since the next element in it is a defocusing quadrupole. Hence the difference in beam distribution between the horizontal and vertical planes. Next, the agreement is quite satisfactory between the measured and simulated profiles. Last, the asymmetric shape of the beam profile MHP44 is due to the low energy inelastic tails of the transmitted protons through TgE. The AHL magnet which bends the beam vertically allows to reveal the existence of such a tail. Nevertheless, it appears that with the present profile measurement, the beam halo is under-represented in MHP41-42. This is under investigation.

\subsection{Proton beam}
\begin{figure*}
\centering 
\includegraphics*[width=8cm]{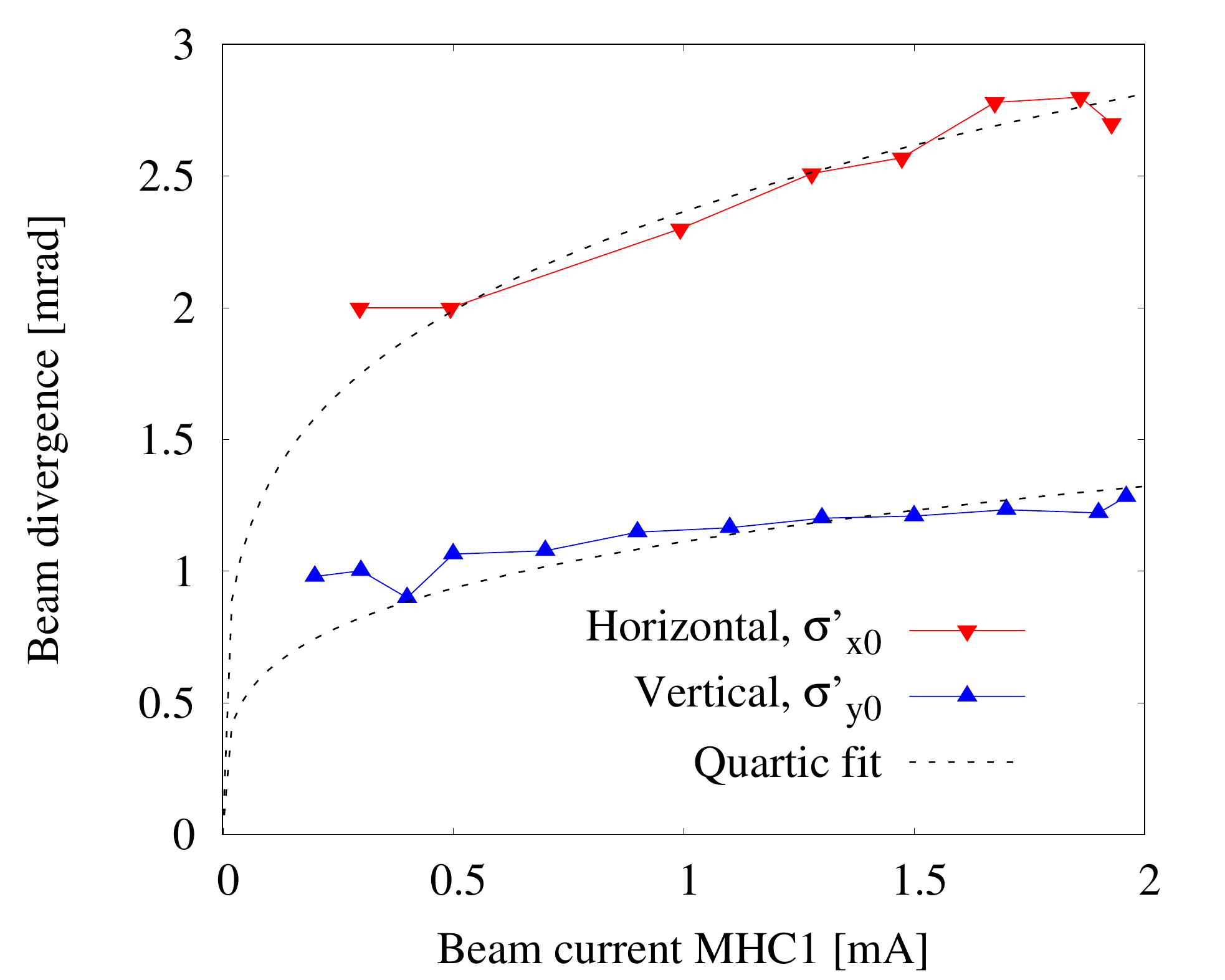}
\caption{Measured divergence of the proton beam impinging on TgE. The profile monitors measurements are repeated for various currents from 0.25 mA to 2 mA. The quartic fit is based on Eq.~(\ref{eq:sigmaxp0}).}
\label{fig:Divergence}
\end{figure*}

Beam studies have shown that the beam properties at the extraction from the main ring change with the beam current due to two fundamental reasons: On one hand, space charge forces act differently on the beam with the current \cite{stammbach2001psi, baartman2013space, kolano2018intensity}. On the other hand, the injected current into the cyclotron chain is modified by cutting into the beam with a moving collimator thereyby changing its emittance. \\
The changing beam conditions from the cyclotron lead to changing beam sizes and divergences at TgE. The latter, denoted $\sigma_0$ and $\sigma{'}_{0}$ are critically important as we shall discuss later on in this paper. A comparison of the measured beam divergences of the proton beam impinging on TgE is shown in Fig.~\ref{fig:Divergence}. Clearly, at beam currents of 0.5 mA and above, the beam divergence can be well fitted with a quartic equation as follows:
\begin{equation}
\sigma{'}_{x0} (I) = \sigma{'}_{x0}(I_{\text{ref}}) \times \left(\dfrac{I}{I_{\text{ref}}} \right)^{1/4}
\label{eq:sigmaxp0}
\end{equation}
and similarly for $y$. In addition, the beam size at TgE can be well approximated by a similar quartic equation \cite{kolano2018intensity} fulfilling $\sigma_{x0}=0.8$ mm and $\sigma_{y0}=1.3$ mm at 2 mA.

\subsection{Probing the losses: Methodology}

\begin{figure}[!h] 
\centering
\hfill
\subfigure{\includegraphics[width=8cm]{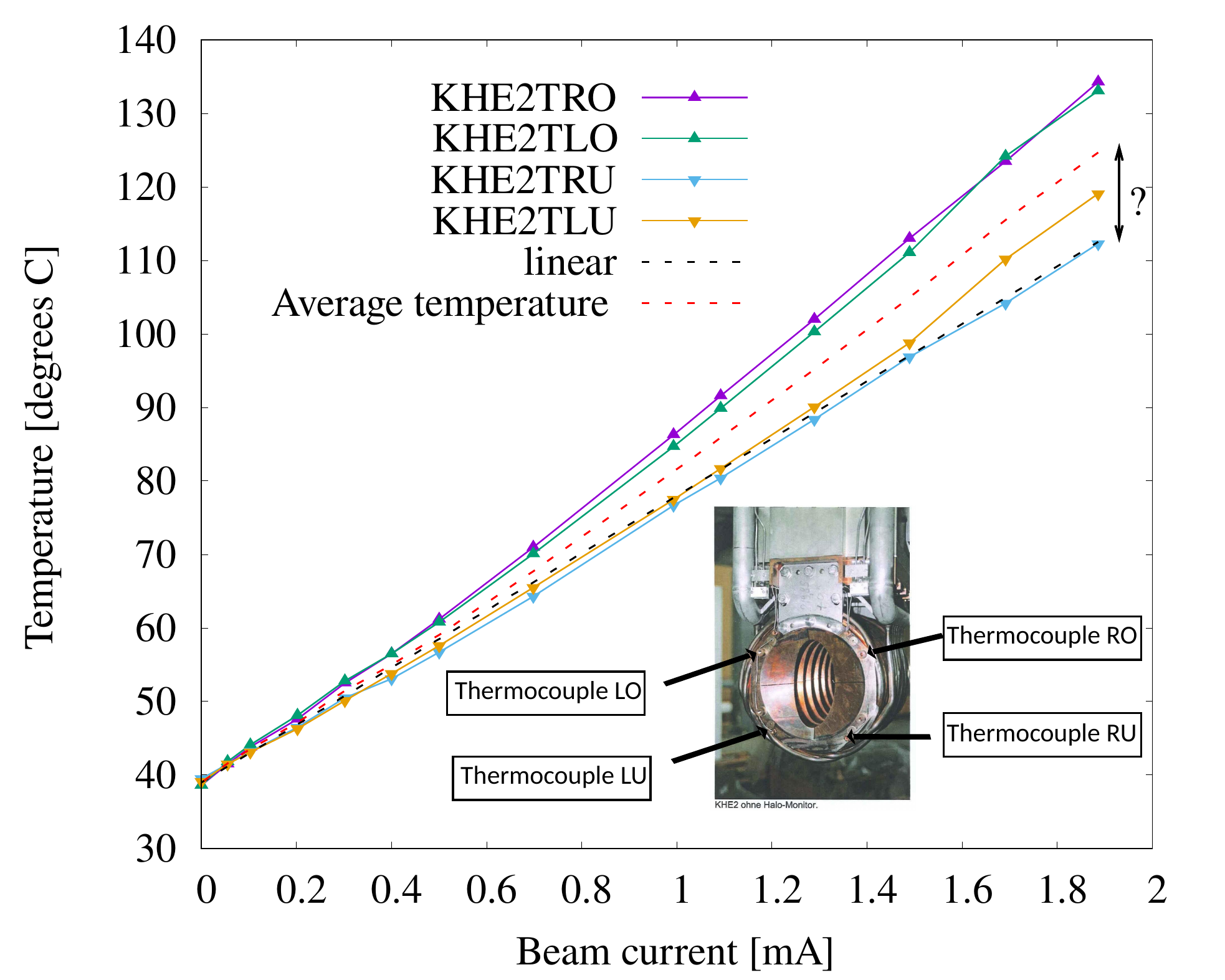}}
\hfill
\subfigure{\includegraphics[width=8cm]{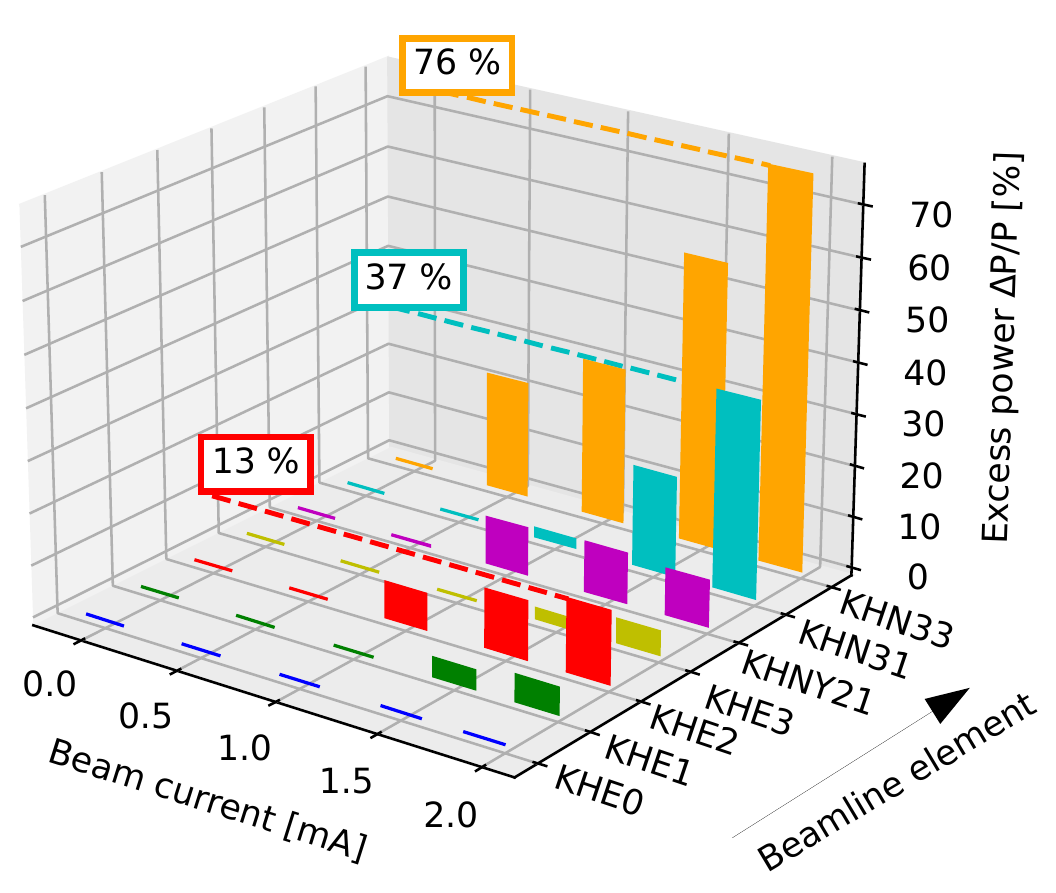}}
\hfill
\caption{Left: Example of measured surface temperature at the entrance of KHE2 collimator as a function of the injected beam current. Right: The histogram of distributed losses $\Delta P/P$ as a function of the beam current is shown for some of the most crucial elements. }
\label{fig:KHE2_surface}
\end{figure}

In our approach, the optics is maintained, and we vary the injected beam current into the cyclotron chain by means of a movable collimator, denoted KIP2 \cite{bib:stetsoncommissioning}. The latter, located on the first turn of the 72 MeV cyclotron, selects the fraction of the remaining proton beam that shall be accelerated and delivered to the different experiments. The beam distribution shall thus evolve with the current, and the distributed beam losses as well as the power depositions shall be intensity-dependent. Following the optics validation, we aim to benchmark the BDSIM developed model of the high power beamline with the power deposition as well as with the beam transmission measurements for various beam currents. However, in order to achieve the latter, it is crucial to understand the provenance of the losses along the beamline.
In an effort to probe their origin, the idea is to inspect the variation of the surface temperature measurements with the current in order to detect the elements whose temperature does not exhibit a linear behavior. The merit of the temperature measurement is that it is an average value so less prone to local showers that may be concealed from the beam loss monitors due to the presence of local shielding or anisotropic shower distribution. Under nominal cooling conditions, and after reaching the steady heat transfer process, we define the relative excess power with respect to the linear behavior as follows:
\begin{eqnarray}
\dfrac{\Delta P}{P} = \dfrac{\mean{T_s}-\mean{T_{s,linear}}}{\mean{T_{s,linear}}-\mean{T_{s0}}}
\end{eqnarray}
where $P$ is the power deposited by the beam in the element, $\mean{T_s}$ is the average surface temperature of the element,~i.e., averaged over all its thermocouples which shall be evenly distributed around the beam optical axis, $\mean{T_{s,linear}}$ is inferred from fitting the measured linear temperature increase at low currents typically below 0.5 mA, and $\mean{T_{s0}}$ is the mean surface temperature at zero current. An increase of $\Delta P/P$ with the current shall be a synonym of increase in relative beam losses. An example of such a measurement averaged over four thermocouples is shown in Fig. \ref{fig:KHE2_surface} for TgE collimator, denoted KHE2, which is the largest source of controlled beam losses in the entire beamline, cutting nearly $10\%$ of the incident beam. The experiment shows an increased power deposition of about $13\%$ in comparison with the linear regime. The same analysis is repeated for all elements equipped with surface temperature measurements. As a summary, the histogram of distributed beam losses is shown in Fig.~\ref{fig:KHE2_surface} for the elements exhibiting important changes: TgE collimators (KHEx), vertical slits KHNY21 inside the AHL magnet, as well as the SINQ collimators (KHN31 and KHN33) which exhibit the largest departure of their power depositions from the linear regime. \\
Nevertheless, one of the major limitations of such a measurement is associated with the thermocouple junction which is subject to embrittlement and corrosion in such harsh environment, thus potentially affecting its calibration and accuracy. 

 Thus, key questions that remain include whether the surface temperature measurement suffers from systematic errors altering the obtained results and to determine the absolute value of the beam power depositions that shall be compared with our BDSIM/GEANT4 Monte Carlo simulations, in particular for the identified elements.
For this reason, our analysis of the losses will be reinforced by comparing with the cooling water temperature measurements (Section~\ref{sec:power_depos}) as well as with the beam transmission measurements (Section~\ref{sec:transmission}). However, before proceeding any further, we shall develop an analytical model to relate the beam losses to the changing beam conditions from the cyclotron in the vicinity of TgE.

\section{Controlled beam losses}

\begin{figure}
\centering 
\includegraphics*[width=10cm]{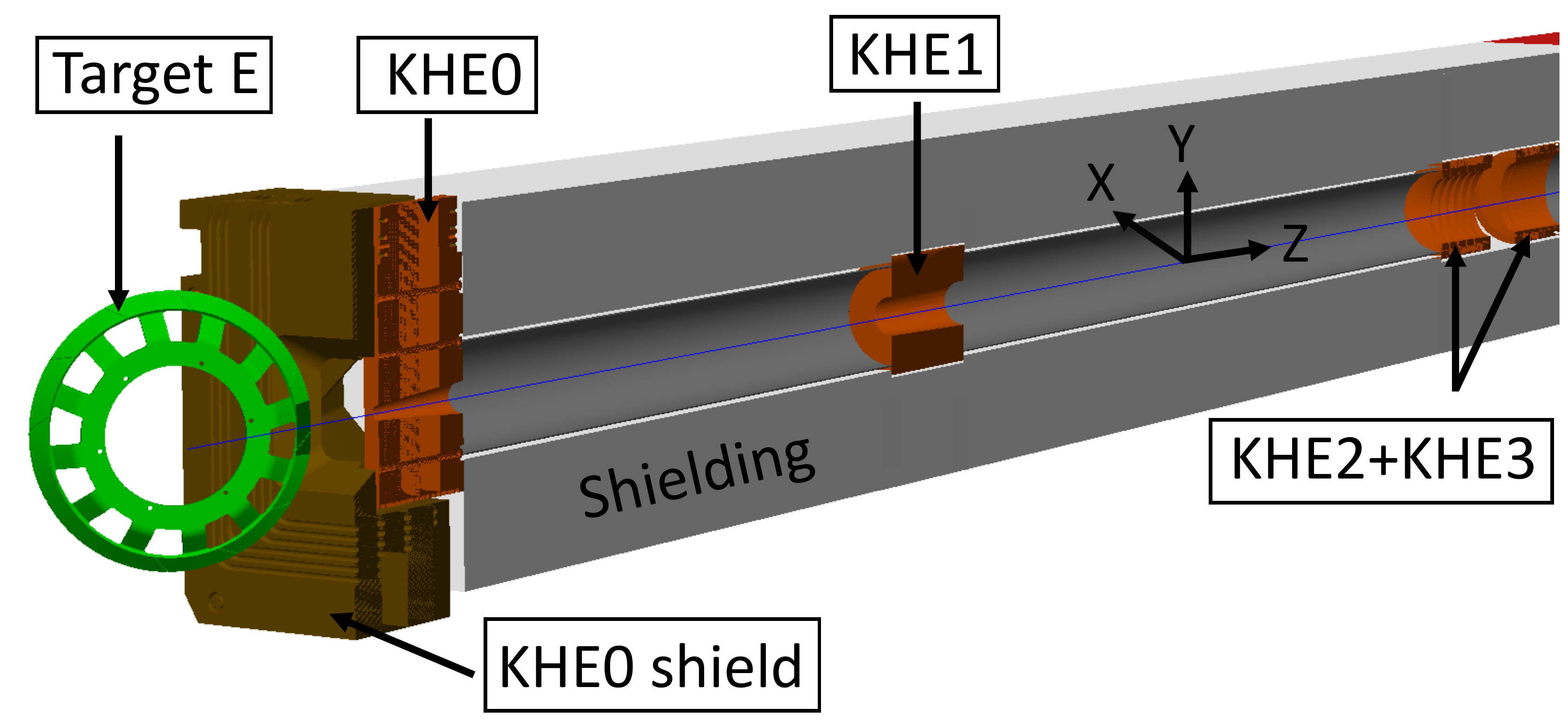}
\caption{BDSIM model of the Target E region (beam impact region in darkgreen). The beam is moving in the longitudinal Z direction. Target E, rotating at 1 Hz, is made of graphite while all collimators are made of Copper. The shielding in which all collimators are enclosed, is displayed in lightgray.}
\label{fig:TgE_coll}
\end{figure}

We first focus our analysis on the losses in the vicinity of TgE. The BDSIM/GEANT4 developed model is shown in Fig.~\ref{fig:TgE_coll} where we can see the target wheel as well as all four subsequent collimators: KHE0 which is acting as a shielding collimator to intercept the secondary showers from TgE, KHE1 which intercepts only a small fraction of the primary beam amounting to $\approx 1.8\%$, KHE2 and KHE3 which, combined, intercept $\approx 13.5\%$ of the highly divergent primary proton beam and exhibit the largest departure of their surface temperatures from the linear regime as shown in Fig.~\ref{fig:KHE2_surface}.
Thus, in order to gain further insights into the origin of the loss increase with the injected beam current from the ring cyclotron, we develop a simple model of the primary beam losses on collimators KHE2 and KHE3 which we will treat as a single block denoted KHE23.
Given that the beam is emittance-dominated, and that the region between TgE and KHE3 is dispersion-free, the first order expansion of the beam size at the entrance of the collimator KHE2 following target E can be written as follows:
\begin{eqnarray}
\sigma_x = \sigma_{x0} + L \left(\sigma{'}_{x0}^2 + \vartheta_{sc}^2 \right)^{1/2} \label{eq:sigmax}
\end{eqnarray}
where $\sigma_{x0}$ is the rms beam size at target E, $L=4.74$ m the distance between target E and the collimator, $\sigma{'}_{x0}$ is the initial beam divergence at the entrance of the target, and $\vartheta_{sc}$ is the rms scattering angle accounting for the multiple Coulomb scattering of the beam at the target. In 1941, Rossi amd Greisen derived the small angle multiple elastic scattering law from Rutherford's single scattering law \cite{rossi1941cosmic}. It results that, for single charged projectile particles, $\vartheta_{sc}$ can be expressed in terms of the radiation length $X_R$ by:
\begin{eqnarray}
\vartheta_{sc}^{R} = \dfrac{E_s}{p \beta c} \sqrt{\dfrac{X}{X_R}} 
\end{eqnarray}
where $X$ is the thickness of the target material, $E_s$ a constant whose value is 14 MeV (21 MeV in the original paper), $p$ the momentum and $\beta c$ the velocity of the incident particle.
Later on, Highland improved the theory \cite{highland1975some} so that 
\begin{eqnarray}
\vartheta_{sc}^{H} = \dfrac{E_s'}{p \beta c} \sqrt{\dfrac{X}{X_R}} \left[1 + 0.038 \ln \left(\dfrac{X}{X_R} \right) \right]
\end{eqnarray}
where $E_s'=13.6$ MeV.
In addition, the radiation length can be approximated by:
\begin{eqnarray}
X_R = 1433 \hspace{1mm} \mathrm{g} \hspace{1mm} \mathrm{cm^{-2}} \dfrac{A}{Z(Z+1)(11.319-\ln Z)} = 42.9745 \hspace{1mm} \mathrm{g} \hspace{1mm} \mathrm{cm^{-2}}
\end{eqnarray}
After impacting TgM, the impinging proton beam energy at TgE reduces to 587 MeV so that $\vartheta_{sc}^{H}=5.83$ mrad. 
Nevertheless, due to the presence of the first two collimators denoted KHE0 and KHE1, the effective scattering angle of the remaining primary proton beam reaching KHE23 is $\vartheta_{sc}=5.06$ mrad as obtained in BDSIM. 

\begin{figure}[!h] 
\centering
\hfill
\subfigure{\includegraphics[width=8cm]{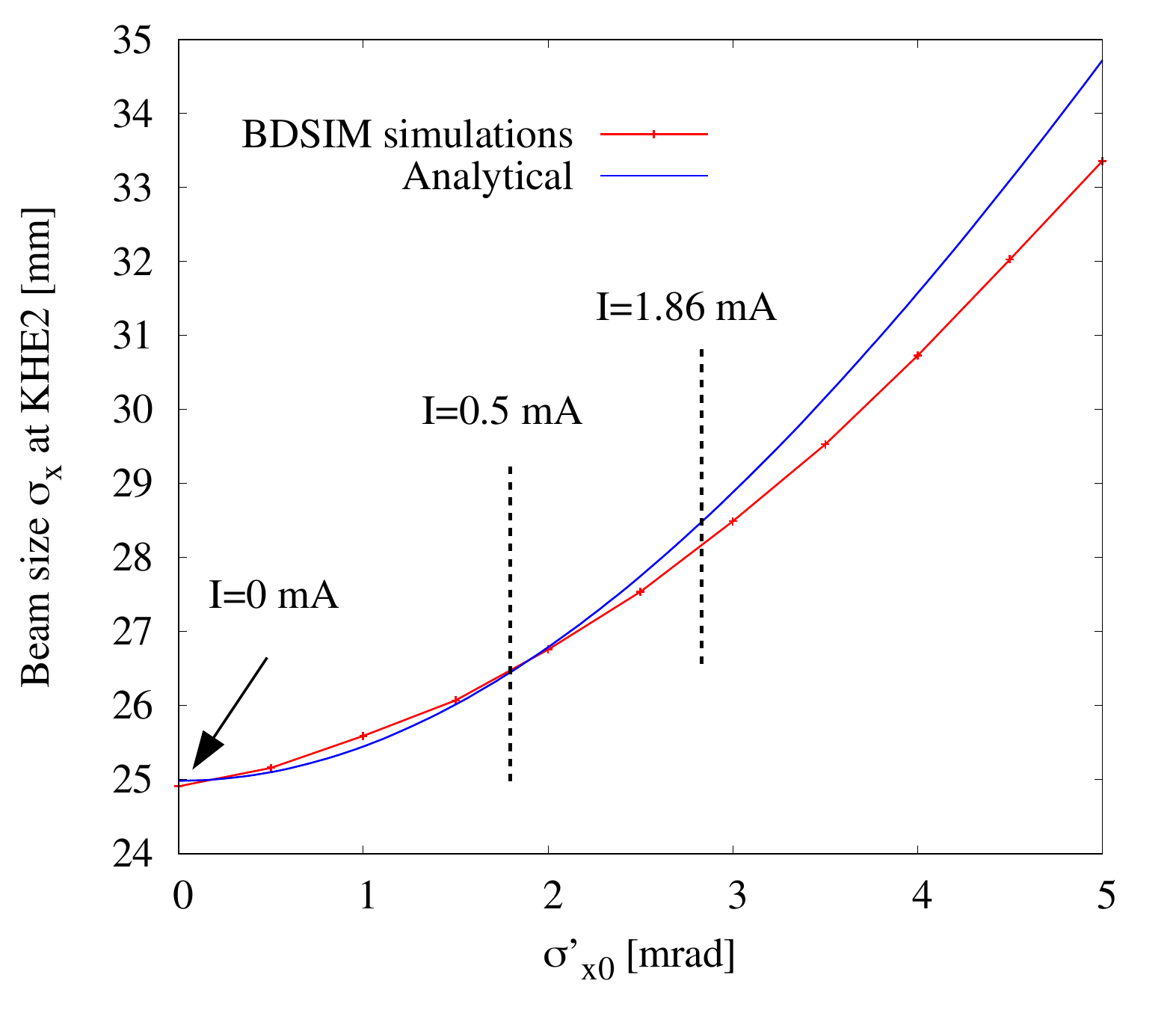}}
\hfill
\subfigure{\includegraphics[width=8cm]{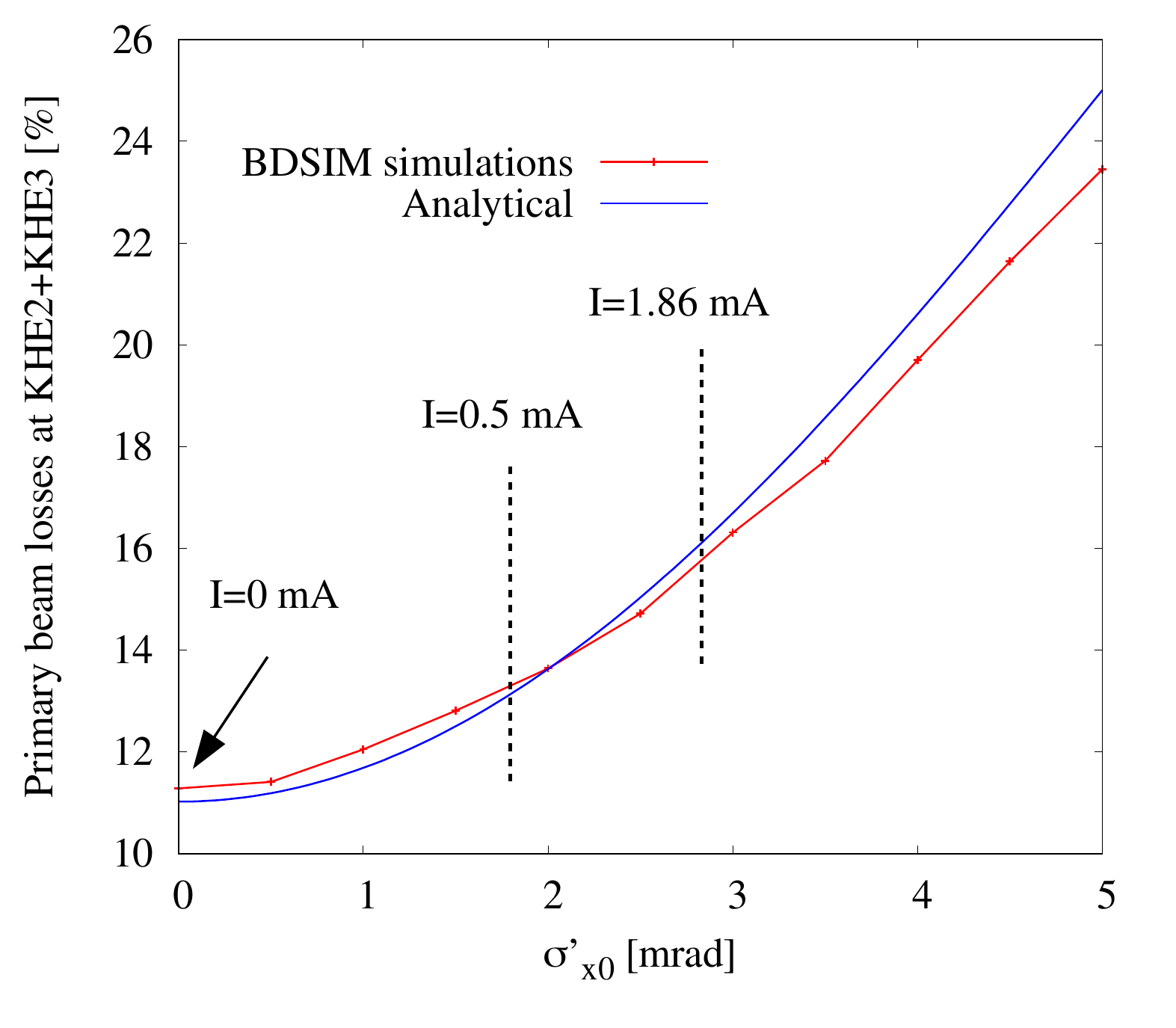}}
\hfill
\caption{Left: Comparison of the simulated beam size at the entrance of KHE2 collimator block with the analytical expression~(\ref{eq:sigmax}) where an initial beam size at target, $\sigma_{x0}=1$ mm, is assumed. Right: Comparison of the simulated fractional primary beam losses at the collimator block with the analytical expression~(\ref{eq:p_KHE23}).}
\label{fig:ploss_KHE23}
\end{figure}

\noindent Given that the horizontal aperture of KHE2 is twice smaller in the horizontal plane than in the vertical one, it is clear that most of the losses take place in the horizontal plane.
Once the horizontal beam size is determined at the entrance of KHE2, the fractional beam losses can be calculated by integrating the fraction of the Gaussian beam which is contained between the values -$a_x$ and +$a_x$ where $a_x=40$ mm is the horizontal aperture at the entrance of KHE2:
\begin{eqnarray}
p_{\text{KHE23}} = \dfrac{I_{\text{KHE23}^-}-I_{\text{KHE23}^+}}{I_{\text{KHE23}^-}} &=& 1 - \dfrac{1}{\sqrt{2\pi} \sigma_x} \int_{-a_x}^{+a_x} \exp \left(-\dfrac{1}{2} \dfrac{x^2}{\sigma_x^2} \right) dx \nonumber \\
&=& \mathrm{erfc} \left(\dfrac{a_x}{\sqrt{2} \sigma_x} \right) = \mathrm{erfc} \left(\dfrac{a_x}{\sqrt{2} \left[\sigma_{x0} + L \left(\sigma{'}_{x0}^2 + \vartheta_{sc}^2 \right)^{1/2} \right]} \right)  \label{eq:p_KHE23} 
\end{eqnarray}
where $\mathrm{erfc}$ is the complementary error function and the subscript +/- denotes the measured beam current at the exit/entrance of the specified element. \\ 
From Eq. (\ref{eq:p_KHE23}), it can be inferred that the initial beam divergence at the target plays a major role in determining the subsequent beam size at the entrance of the collimator and consequently the primary losses therein. In the ideal case, the beam divergence at the target shall be vanishing. Nevertheless, due to intensity-dependent effects, it increases with the current according to a quartic law given by Eq. (\ref{eq:sigmaxp0}). In order to investigate the impact of such a quantity, we vary the initial beam divergence at the target and compute the primary beam losses by means of Eqs.~(\ref{eq:sigmaxp0}) and (\ref{eq:p_KHE23}). As illustrated in Fig.~\ref{fig:ploss_KHE23}, due to the smallness of the beam size on target, the increased divergence of the proton beam impinging on TgE plays the dominant role to explain the increased losses on subsequent collimators, particularly on KHE23. This will be discussed in detail in the next sections.

\section{Power depositions} \label{sec:power_depos}
In what follows, we discuss two approaches to measure the deposited power along the beamline: one that relies on the water temperature measurements of the cooling circuit of the elements, and the second one relies on the surface temperature measurements of the element. Although the first approach is applicable for actively-cooled elements only, the second one is utilized to infer the effective heat transfer coefficient and is more generally applicable to components fulfilling the criterion for lumped system analysis.

%\subsection{Benchmarking BDSIM/MCNP simulations}

%Collimators are used to form the limiting apertures that shall cut the largely scattered particles and prevent them from depositing their energy downstream. In order to validate the accuracy of the model, we perform a comparison of MCNP6 versus BDSIM: the idea is to compare the scattering angles as well as the deposited powrs at several stages of the collimation process.

\subsection{Power depositions for actively-cooled elements}

\begin{figure}[!h] 
\centering
\hfill
\subfigure{\includegraphics[width=8cm]{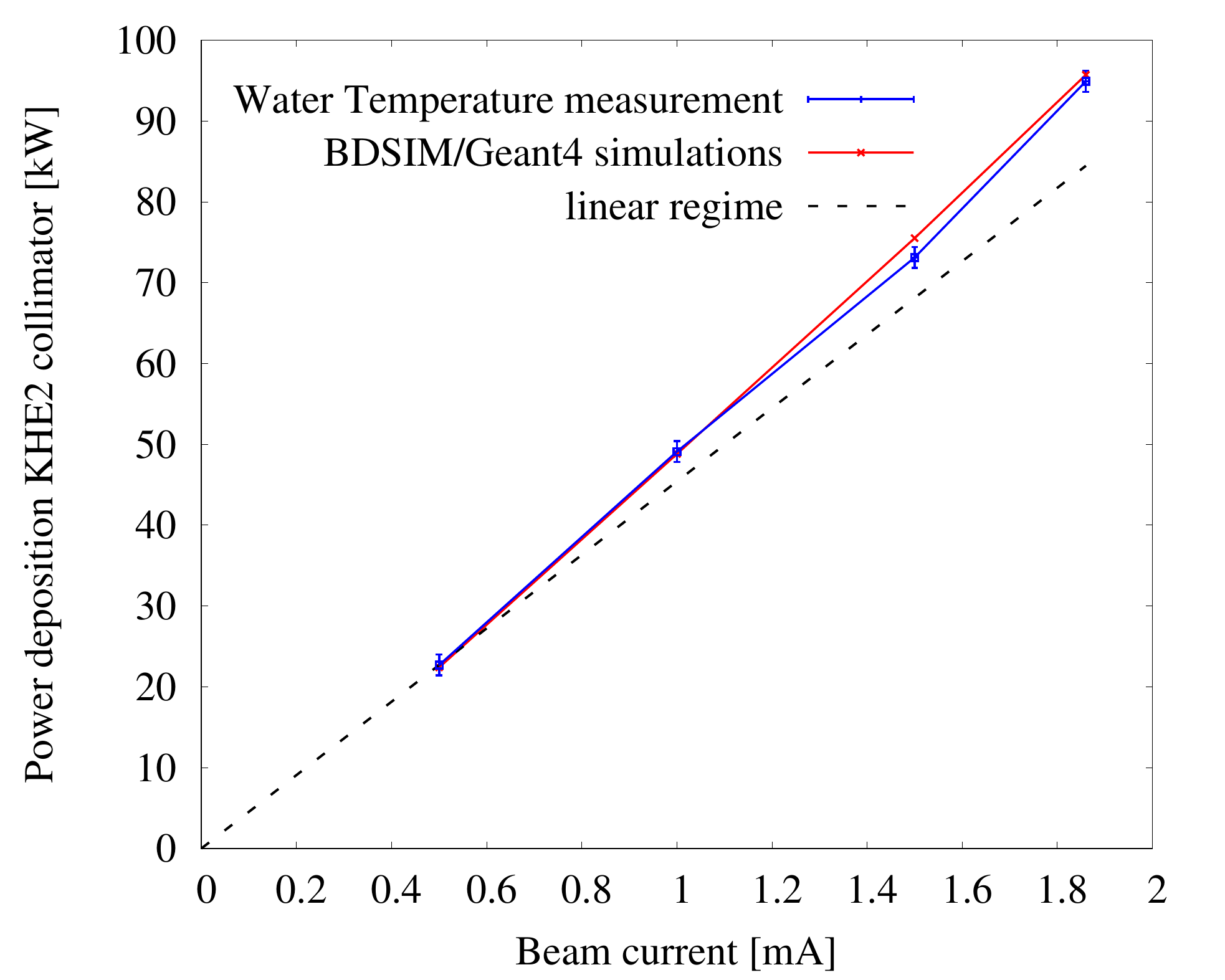}}
\hfill
\subfigure{\includegraphics[width=8cm]{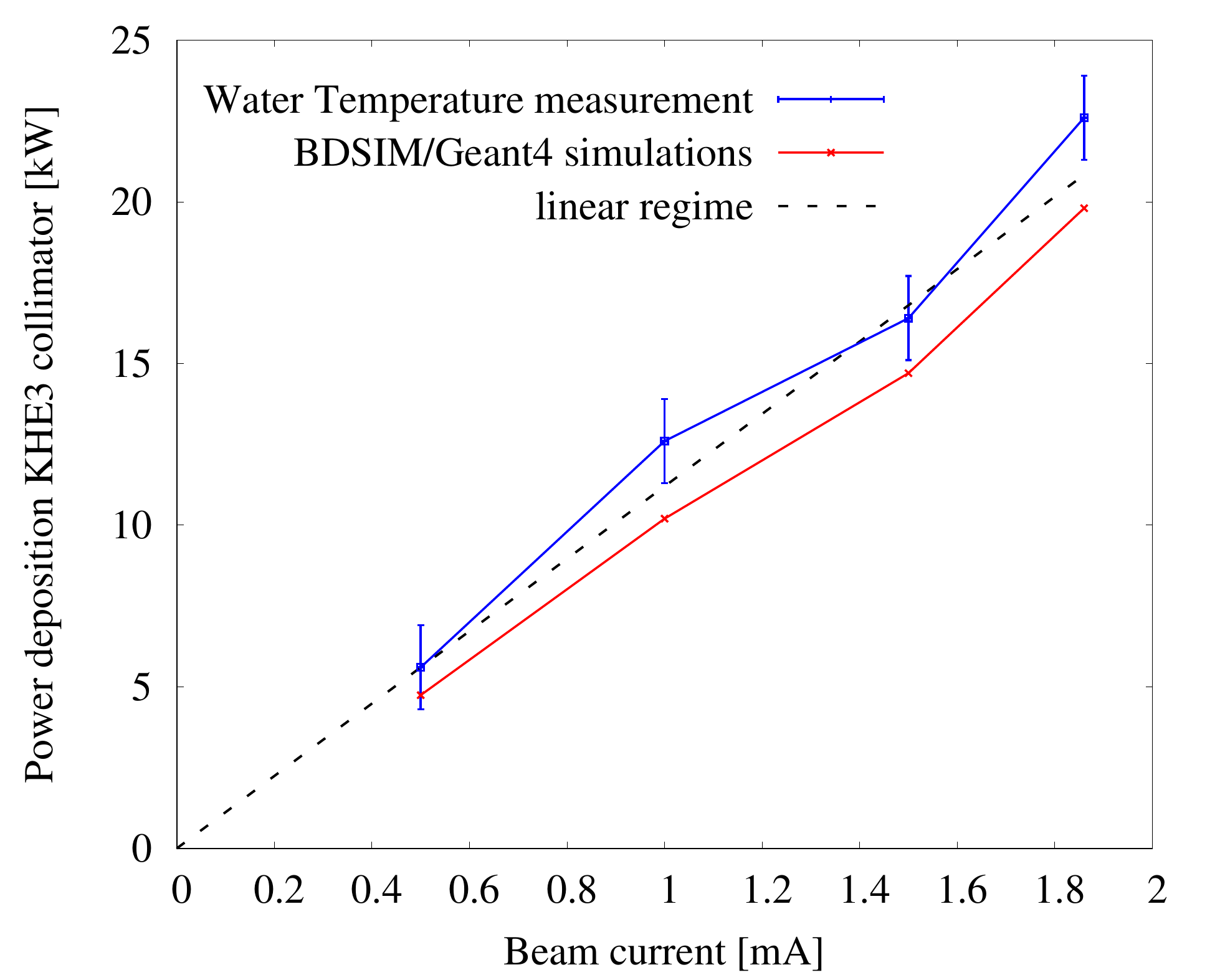}}
\hfill
\caption{Measured power deposition as a function of the injected beam current and comparison with BDSIM/Geant4 simulations.}
\label{fig:P_depo_KHE23}
\end{figure}

Under steady operating conditions, i.e., CW beam from the cyclotron without any beam trips, and steady flow conditions, the rate of heat transfer out of the cooled system, which is equivalent to the amount of beam power deposited within its volume can be written as follows
\begin{eqnarray}
P = \dot{m} C_p \Delta T
\end{eqnarray}
where $\dot{m}=dm/dt$ denotes the mass flow rate in kg/s,~i.e., the amount of mass flowing through the cross section of the cooling pipe, $C_p$ is the isobaric specific heat capacity of the coolant in units of J/(kg.K), and $\Delta T$ is the temperature difference between the outlet and inlet water (inlet temperature can also be inferred from zero beam current measurement). For water, $C_p$ is almost independent of the temperature such as $C_p = 4.2$ kJ/(kg.K) for all our measurements. In steady operation, there is no change in the temperature of the object with time at any point. In addition, the entrant/exiting fluid temperatures shall remain constant. \\
The beam current was varied from 0 to 1.86 mA by step sizes of $\sim$ 0.5 mA and the time between two consecutive measurements is approximately 1300 s with the constraint that no beam trip shall occur during this process. The measured power depositions at KHE2 and KHE3 collimators are shown in Fig.~\ref{fig:P_depo_KHE23} and compared to the simulation results. The latter incorporated the changing beam conditions at TgE as obtained from the beam profile measurements and accurately describe the observed excess power with respect to the linear regime. In particular, it is shown that the deposited power in KHE2 at 1.86~mA is larger by 11~kW in comparison with the linear regime. This represents a 12\% increase which is in excellent agreement with the analysis of the surface temperature measurements (see Fig.~\ref{fig:KHE2_surface}).
A full comparison of the power deposition measurements in all TgE collimators (depicted in Fig. \ref{fig:TgE_coll}) is finally shown in Table~\ref{Tab:power_depos_steady_state}. For all power depositions, the agreement between measurement and simulations is better than 15\%.  Summing over all TgE collimators, the simulations are less than 1\% higher than  the measurements. This allows to further establish the validity of the built-in model of the beamline, the initial beam conditions as well as the physics processes which are mainly driven by the scattering processes.
\begin{table}[htb]
\centering
\caption{Power deposition measurements for an injected beam current of 1.86 mA and comparison with BDSIM simulations\label{Tab:power_depos_steady_state}}
\def\arraystretch{1.2}
\begin{tabular}{lllll}
\hline
\hline
\multicolumn{1}{l} {Target E}&\multicolumn{1}{l}{Mass flow}&\multicolumn{1}{l}{Water temperature}&\multicolumn{1}{l}{Power deposition} &\multicolumn{1}{l}{Power deposition} \\
\multicolumn{1}{l} {collimators}&\multicolumn{1}{l}{rate [kg/s]}&\multicolumn{1}{l}{difference [K]}&\multicolumn{1}{l}{measurement [kW]}&\multicolumn{1}{l}{from simulations [kW]} \\
\hline
KHE0&1.2&4.1 $\pm$ 0.1&20.7 $\pm$ 0.5&23.3\\

KHE0 shield&1.2&3.9 $\pm$ 0.1&19.7 $\pm$ 0.5&21.9\\

KHE1&1.2&3.0 $\pm$ 0.1&15.1 $\pm$ 0.5&15.3\\

KHE2&3.0&7.5 $\pm$ 0.1&94.9 $\pm$ 1.3&94.4\\

KHE3&3.0&1.8 $\pm$ 0.1&22.6 $\pm$ 1.3&19.1\\

Total&-&-&173.0 $\pm$ 4.1&174.0\\
\hline
\hline
\end{tabular}
\end{table}
Unfortunately, not all installed beamline elements feature water temperature control. Instead surface temperature measurements are more commonly used to check that the elements remain within a safety margin from their melting temperatures and have a faster response time, although not fast enough to be used as an interlock. 

\subsection{Effective heat transfer coefficient}
\begin{figure}[!h] 
\centering
\hfill
\subfigure{\includegraphics[width=8cm]{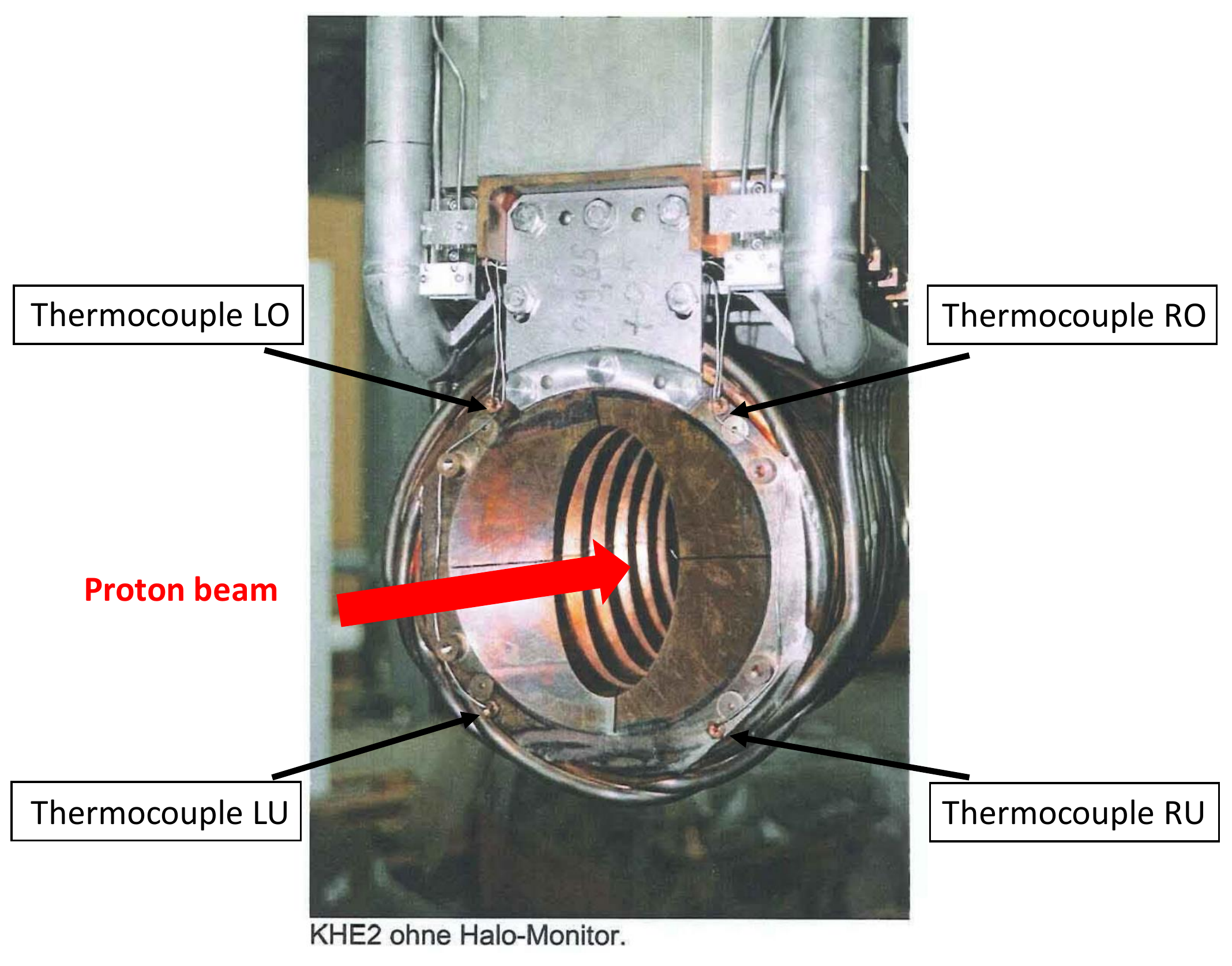}}
\hfill
\subfigure{\includegraphics[width=8cm]{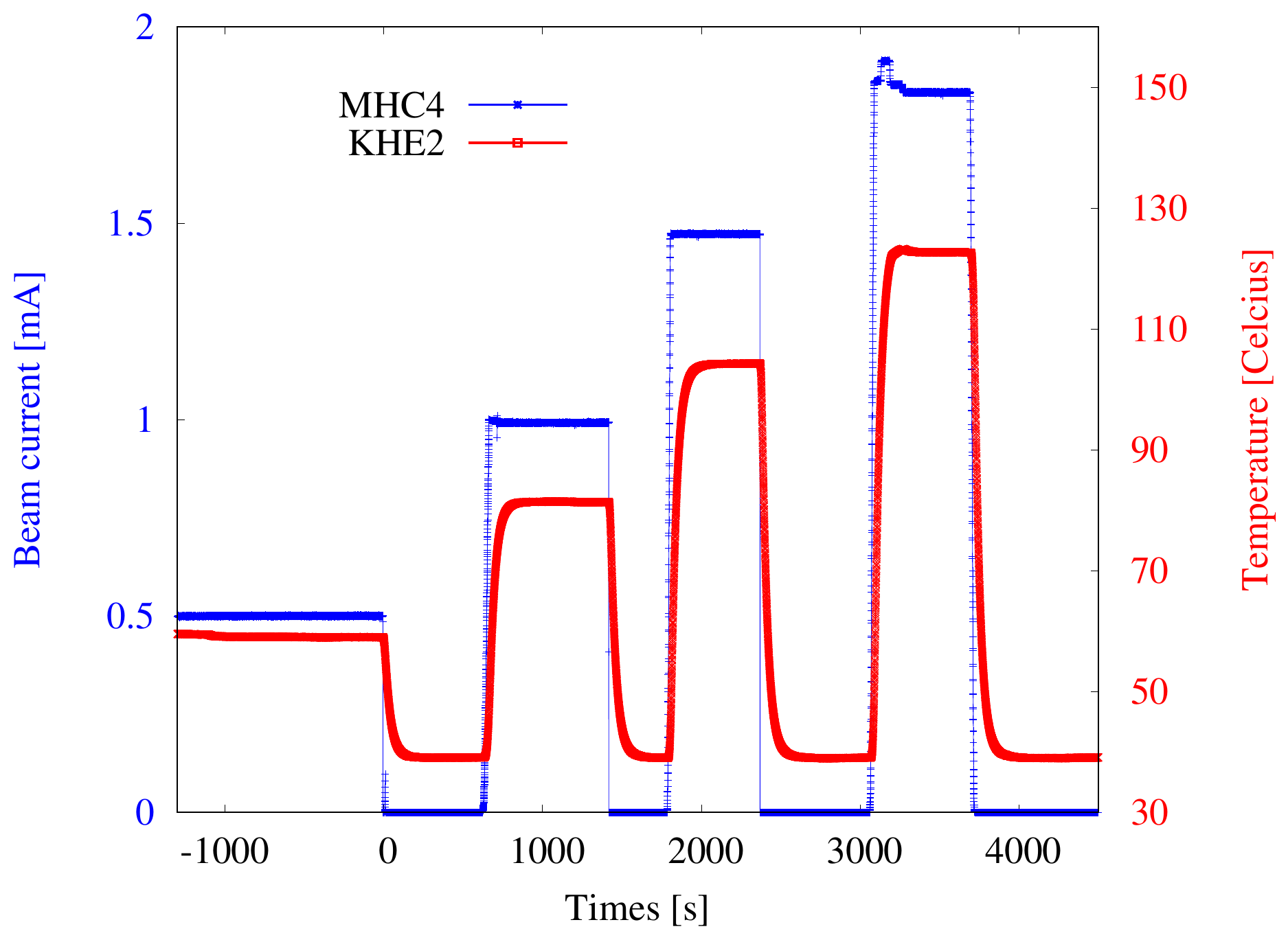}}
\hfill
\caption{Left: KHE2 collimator with its complex teeth structure. Right: Measured temperature profile for various currents from 0.5 mA to 1.86 mA. }
\label{fig:thermocouples_exp}
\end{figure}

According to Newton's law of cooling, the heat transfer coefficient $h$ establishes a linear correlation of the rate of heat flux out of the body and the temperature difference between the body surface and its surroundings:
\begin{eqnarray}
P = h A_s \left(T_s - T_{\infty}\right) \label{eq:heat_transfer_coeff}
\end{eqnarray}
where $T_s$ is the surface temperature of the object in steady operation, and $T_{\infty}$ is the temperature of the surrounding medium.
It is important to note that $h$ is not a property of the fluid but rather an experimental quantity that depends on all parameters influencing the heat transfer process such as the material composition, its surface geometry, the fluid velocity and its properties as well as the quality of the brazing, i.e., the thermal connection between the cooling tubes and the body (copper for instance). \\
Given that the temperature change of all elements remains below 100 degrees, the heat load by radiation is discarded (in order to approach the kW sensitivity level of the measurement, the body temperatures shall exceed 1000 degrees). Thus, in our analysis, we assume all the heat loss to occur by convection. The aim is to devise a beam-based experimental method to determine the effective heat transfer coefficient $h A_s$. First, we focus our analysis on the elements which are actively-cooled so that the beam deposited power is directly determined as shown in the previous section: By means of controlled beam interruptions at various currents, the surface temperature of the object and the surrounding medium is determined. The temperature is the average signal from four thermocouples that are evenly distributed around the optical axis as shown in Fig. \ref{fig:thermocouples_exp}. By means of Eq.~(\ref{eq:heat_transfer_coeff}), the effective heat transfer coefficient is determined for a specific current and the result summarized in Table \ref{Tab:heat_transfer_coeff}. Performing the measurement for several currents allows to verify the constancy of the calculated coefficient. In addition, we establish that the uncertainty of the calculated quantities is mainly limited by the uncertainty of the water temperature measurement. This reinforces the validity of both water and surface temperature measurements as an effective way to determine the beam deposited power and allows to verify the accuracy of the required calibration of the thermocouple junctions.

\begin{table}[htb]
\centering
\caption{Beam-based determination of the integrated heat transfer coefficient for an injected beam current of 1.86 mA\label{Tab:heat_transfer_coeff}}
\def\arraystretch{1.2}
\begin{tabular}{lllll}
\hline \hline
\multicolumn{1}{l} {Beamline}&\multicolumn{1}{l}{Mass}&\multicolumn{1}{l}{Surface temperature}&\multicolumn{1}{l}{Heat transfer} \\
\multicolumn{1}{l} {element}&\multicolumn{1}{l}{$\rho V$ [kg]}&\multicolumn{1}{l}{difference $\left(T_s - T_{\infty} \right)$ [K]}&\multicolumn{1}{l}{coefficient $h A_s$ [kW/K]} \\
\hline
KHE0&120&21.1 $\pm$ 0.5&0.981 $\pm$ 0.048\\

KHE1&105& 17.6 $\pm$ 0.5& 0.858 $\pm$ 0.054\\

KHE2&88.6&83.7 $\pm$ 0.5&1.134 $\pm$ 0.022\\

KHE3&82.7&24.8 $\pm$ 0.5&0.911 $\pm$ 0.072\\

\hline \hline
\end{tabular}
\end{table}
A second approach which is general to all elements, i.e., actively-cooled or not, is based on measuring the transient behavior of the surface temperature profile after an abrupt interruption of the proton beam. The measurement was conducted for various steady state regimes reached at different beam currents as illustrated in Fig.~\ref{fig:thermocouples_exp}. The core idea is to fit the measured transient decay with the theoretical solution in order to infer the time constant of such a decay. Given that such a theoretical solution is dependent on the geometry of the targeted element, we rely on a simplified approach based on the lumped system analysis to solve the problem: according to this model, the object (collimator, slits, dipole, etc) is characterized by a uniform temperature at all times during the heat transfer process. It results that its surface temperature $T(t)$ shall drop exponentially according to:

\begin{eqnarray}
\dfrac{T(t)-T_{\infty}}{T_i - T_{\infty}} = \exp \left(-\dfrac{t}{\tau} \right)
\end{eqnarray}
where $T_i$ is the initial surface temperature of the element, i.e., at steady state regime, $T_{\infty}$ is the final surface temperature of the element, i.e., at zero beam current, and $\tau$ is the time constant of the exponential decay which is given by
\begin{eqnarray}
\tau = \dfrac{\rho V C_p}{hA_s} \label{eq:tau}
\end{eqnarray}
where $\rho$ and $V$ denote the density and volume of the element exchanging the heat, while $C_p$ is the specific heat capacity of its material. In particular, all collimators are made of Copper for which $C_p$ obeys a linear function of the temperature above 270 K and is well fitted by \cite{bib:banerjee2005evaluation}
\begin{eqnarray}
C_{p,cop} \left[\dfrac{J}{kg.K} \right] = 358.4 + 0.1009 T \text{ for T} \geq 270 K
\end{eqnarray}
Fitting the exponential temperature decay yields $\tau$ and consequently $h A_s$. Furthermore, establishing that the transient decay is independent of the beam current is crucial for the analysis and is a requirement to verify the calibration of the thermocouple junction.
One way to fulfill the conditions of the lumped system analysis is to switch off the cooling circuit and inject a low current beam to slowly reach the uniform temperature profile within the element. After this, the water flow shall be restored and the beam interrupted. However, this will not be pursued here. 
Last but not least, the beam halo scraper KHNY21 is one of the actively-cooled elements that is not equipped with water temperature control yet playing an important role to explain the losses. Given the relatively low water flow rate of 0.2~kg/s for such an element, we rely on the lumped system analysis to estimate its effective heat transfer. This yielded a deposited power therein of roughly 10~kW for an injected beam current of 1.86~mA while our simulations indicate a heat load close to the 8~kW level. As will be discussed in the next section, more dedicated effort is needed in order to determine the precise opening of such beam halo scrapers which determines the beam losses therein.

\subsection{Outcomes}

The power deposition measurements confirmed the BDSIM/GEANT4 Monte Carlo simulations which showed that the main reason for the transmission drop with the beam current is the increase in the primary proton beam losses at TgE collimators. The large beam cut-off at such collimators, amounting to nearly 17$\%$ of the primary proton beam, is driven by the requirement to achieve low loss transport to the subsequent beamline elements, in particular where hands-on maintenance is allowed, and deliver the beam with the appropriate conditions to SINQ target. A condition sine qua non to reach such a good agreement between the power deposition measurements and the Monte Carlo simulations is the implementation of the initial beam conditions from the beam profile measurements at different injected currents from the main ring cyclotron. In particular, we showed that KHE2 plays the most prominent role to explain the losses and its power deposition does not scale linearly with the current due to the changing beam size and more importantly the changing beam divergence at the target. \\
Two approaches were discussed to determine the deposited power within the beamline elements. Although the water temperature measurement is a direct approach, it is clear that the surface temperature measurement, which is the fastest approach can be of interest for such elements that are not actively-cooled or to probe the localized beam losses within specific volumes. Combining both approaches enabled a direct beam-based determination of the heat transfer coefficient for the most prominent beamline components. \\
Furthermore, with the aim to maintain HIPA at the forefront of the intensity frontier, considerations are on the way to increase the nominal beam current from 2.2 mA to 3 mA \cite{grillenberger2021high, seidel2007upgrade}. Extrapolating to 3 mA yields a horizontal beam divergence on target of $\sim$ 3.1 mrad so that the deposited power on KHE2 shall exceed 150 kW. Under such a condition, the collimator will exceed the maximum allowed heating load. To avoid the risk of thermomechanical failures, a new system was designed and constructed that shall overcome such a limitation \cite{lee2010new}. Based on the successful benchmarking of the power deposition with the measurements for the existing collimation system, we perform a one-to-one comparison of the impact of replacing KHE23 block with the newly designed version on the subsequent power deposition. The results are summarized in the appendix \ref{app:A} and serve as a validation test of the already optimized collimator design for 3 mA operation.

\section{Beam transmission} \label{sec:transmission}
Measuring the staged as well as the final beam transmission is crucial to gain better insights into the distributed beam losses along the SINQ beamline and verify the accuracy of the developed model. 

\subsection{Intensity-dependent beam transmission}

\begin{figure}
\centering 
\includegraphics*[width=10cm]{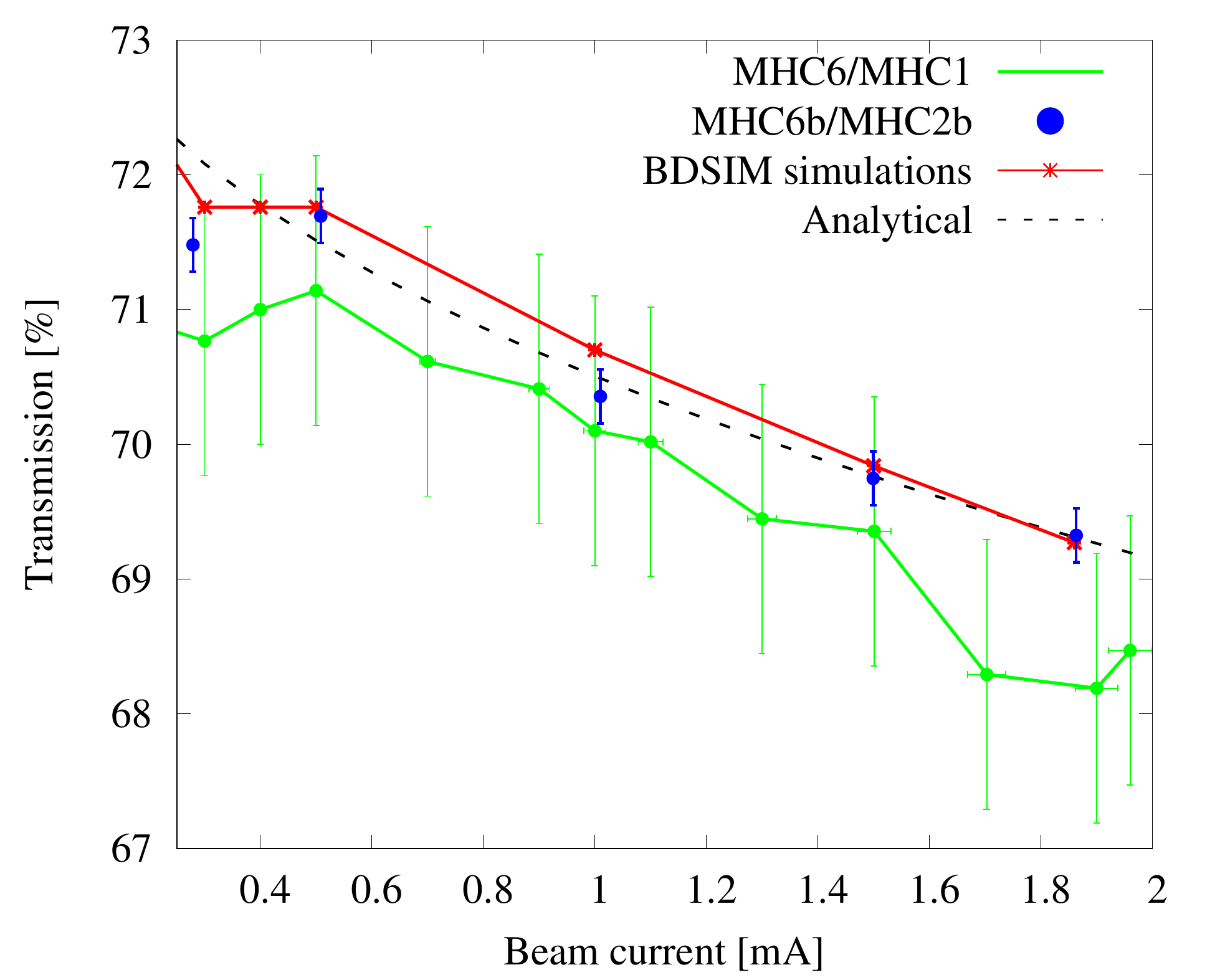}
\caption{Measured beam current transmission for various extracted currents from the ring cyclotron (MHC1) and comparison with the analytical as well as the simulation results. The absolute current measurements (MHC6b/MHC2b) were performed after controlled beam interruptions for recalibration purposes.}
\label{fig:transmission}
\end{figure}

Next, we shall probe the effectiveness of the developed model to reproduce the dependence of the beam transmission on the injected beam current.
As was established earlier, the main source of the increasing power deposition with the current in the TgE region is the KHE23 collimator block. In order to better clarify this, we make the assumption that the start-to-end beam transmission drop with the current up to MHC6b (see Fig. \ref{fig:3D_model}) can be fully explained by the transmission drop inside the collimator block KHE23. This can be formulated as follows:

\begin{eqnarray}
R_{\text{MHC6b}} = \dfrac{I_{\text{MHC6b}}}{I_{\text{MHC2b}}} &=& \dfrac{I_{\text{KHE23}^-}}{I_\text{MHC2b}} \times \dfrac{I_{\text{KHE23}^+}}{I_{\text{KHE23}^-}} \times \dfrac{I_\text{MHC6b}}{I_{\text{KHE23}^+}} \nonumber \\
  &=& R_{\text{KHE23}^-} \times \mathrm{erf} \left(\dfrac{a_x}{\sqrt{2} \left[\sigma_{x0} + L \left(\sigma{'}_{x0}^2 + \vartheta_{sc}^2 \right)^{1/2} \right]} \right) \times R_{\text{KHE23}^+} 
\label{eq:transmission_MHC6b}
\end{eqnarray}
where we made use of Eq. (\ref{eq:p_KHE23}) and where we assume that both $R_{\text{KHE23}^-}$ and $R_{\text{KHE23}^+}$ are constants, i.e., independent of the beam current. The beam initial conditions at TgE are obtained from the profile measurements as shown in Fig.~\ref{fig:Divergence}. A comparison of the transmission measurements relying on the absolute current monitors (Bergoz monitors MHC2b and MHC6b) on one hand and the standard resonators (MHC1, MHC6) on the other hand \cite{duperrex2010current} is shown in Fig.~\ref{fig:transmission}. (MHC1, MHC2b) are located upstream of TgM, while (MHC6, MHC6b) are located downstream of TgE~\cite{reggiani2018improving}. At no surprise, the BDSIM simulations which relied on the profile measurements to infer the incoming beam conditions towards TgE agree with the current measurements, to within 0.4$\%$ for all currents. The quality of the analytical formula (\ref{eq:transmission_MHC6b}) to fit the current measurements above 0.5 mA suggests the model is valid in that region. This is expected, given the quality of the quartic fit of the beam divergence above 0.5 mA (see Fig.~\ref{fig:Divergence}). This also implies that almost all losses in the  beamline up to the Bergoz monitor MHC6b are controlled beam losses. Such a result enables us to determine $R_{\text{KHE23}^-}$ and $R_{\text{KHE23}^+}$:  Owing to the very good overall agreement between the measured and simulated power deposition calculations (see Table~\ref{Tab:power_depos_steady_state}), the BDSIM simulations are used to determine the equivalent primary beam losses up to the entrance of KHE2. This yields $R_{\text{KHE23}^-}=0.8595$. From the aforementioned fit, it results that $R_{\text{KHE23}^+}=0.9573$. Thus, of the remaining primary proton beam exiting KHE3, $4.27\%$ is deposited in the region ]KHE3, MHC6b]. In addition, at low currents where a pencil beam can be assumed, a saturation effect is observed and the losses are nearly insensitive to the initial beam conditions. This is due to the dominant effect of the primary proton beam interaction at TgM which increases the divergence of the beam reaching TgE. This is further supported by our analysis of the relative excess power which showed that the losses are nearly unchanged for beam currents below 0.5 mA. \\
Last but not least, it is important to note that achieving such a good agreement between the simulated and measured beam transmission required to determine the opening of KHNY21 slits (and to a lesser degree KHNX22 slits). Such beam halo scrapers are heuristically optimized to reduce the losses downstream. Benchmarking the optics is not sufficient to infer their precise opening and given the lack of water temperature measurements for such element, one relied on benchmarking the power deposition from BDSIM simulations with the values inferred from the surface temperature measurement approach as discussed earlier: the measured losses at KHNY21 indicate that the latter cuts 1.3 $\%$ of the injected beam which is about 0.4 $\%$ larger than the simulated values hence the uncertainty of the overall simulated transmission. \\
In order to break into the precision level of the absolute current monitors, which is of the order of $0.2\%$ \cite{Priv_Comm_PA}, it is clear that a more dedicated effort is needed to accurately determine the opening of the beam halo scrapers. In addition, more beamline elements shall be equipped with water and/or surface temperature measurements in order to determine their power depositions. The latter shall be utilized as a precise benchmarking tool of the optics and physics processes as already demonstrated in the previous sections.

\subsection{SINQ transmission}
Although the beam intensity is a crucial observable to assess the losses, the effective beam reaching the target can be substantially different if the losses occur at the very last stage of its delivery, i.e., where no beam current measurement is available. Such a scenario is likely to occur given the need to protect the high power spallation target against overfocused beam. 
For instance, the analysis of the surface temperature measurement revealed that the SINQ collimators represent an important source of beam losses (see Fig. \ref{fig:KHE2_surface}). In order to elucidate their impact on the final beam transmission, and given the lack of a beam current monitor right before SINQ target, a dedicated experiment was performed where the injected beam current was varied and the deposited beam power measured for all 3 KHN collimators that we denote $P_{\text{KHN}}$. These are connected and cooled in series so that measuring the water temperature difference as discussed earlier shall yield the amount of deposited power within. In addition, treating all 3 KHN collimators as a single block, and taking the average temperature from all 12 thermocouples (4 for each), the surface temperature measurement approach for an injected 1.86 mA beam yielded an effective heat transfer coefficient $h A_s = 0.891$ kW/K. The consistency of both measurements is validated for various beam currents as shown in Table~\ref{Tab:power_depos_KHN}. In consequence, the SINQ transmission shall be expressed as follows
\begin{eqnarray}
R_{\text{SINQ}} = \dfrac{I_{\text{SINQ}}}{I_{\text{MHC2b}}} &=& R_{\text{MHC6b}} \times \dfrac{I_\text{SINQ}}{I_{\text{MHC6b}}} \nonumber \\
&=& R_{\text{MHC6b}} \times \left[1-\dfrac{P_\text{loss}}{P_\text{beam}} \right] \nonumber \\
&=& R_{\text{MHC6b}} \times \left[1-\dfrac{\left(1+\alpha \right) P_\text{KHN}}{I_{\text{MHC6b}} \times E_p} \right]
\label{eq:transmission_SINQ}
\end{eqnarray}
where $E_p= 570$ MeV is the incoming proton beam energy to SINQ and the factor $\alpha$ ($\alpha > 0$) represents the excess power deposited on the shielding surrounding KHN. This was estimated by tweaking the focusing strength of the last quadrupole doublet before SINQ, and comparing the power increase at SINQ to the power decrease within the collimators. Note that the measured power deposition at SINQ represents $\sim 64\%$ of the incoming beam power (only a fraction of the incoming proton beam energy is deposited at SINQ). The uncertainty is such that $\alpha$ lies in the range [0.4:1].
This was obtained by trying out several optics where the beam waist preceding SINQ is slightly displaced and the power deposited in the shielding computed. The important uncertainty is due to the lack of enough profile measurements in the vicinity of SINQ as well as the fact that the beam conditions are intensity-dependent. The results are finally summarized in Table \ref{Tab:power_depos_KHN}.

\begin{table}[htb]
\centering
\caption{Power deposition measurements for SINQ collimators and deduced transmissions \label{Tab:power_depos_KHN}}
\def\arraystretch{1.2}
\begin{tabular}{lllll}
\hline \hline
\multicolumn{1}{l} {Beam current [mA]}&\multicolumn{1}{l}{$P_{\text{KHN}}$ [kW] from}&\multicolumn{1}{l}{$P_{\text{KHN}}$ [kW] from} &\multicolumn{1}{l}{$R_{\text{MHC6b}} [\%]$}&\multicolumn{1}{l}{$R_{\text{SINQ}} [\%]$}\\
\multicolumn{1}{l} {at MHC2b}&\multicolumn{1}{l}{water temperature}&\multicolumn{1}{l}{surface temperature}&\multicolumn{1}{l}{}&\multicolumn{1}{l}{}\\
\hline
0.5&7.2&7.5&71.7 $\pm$ 0.2 &67.4 $\pm$ 0.9\\

1.0&19.8&19.6&70.4 $\pm$ 0.2 &64.5 $\pm$ 1.2\\

1.5&32.0&32.0&69.7 $\pm$ 0.2 &63.3 $\pm$ 1.3\\

1.86&44.7&44.7 (ref)&69.3 $\pm$ 0.2 &62.1 $\pm$ 1.4\\
\hline \hline
\end{tabular}
\end{table}
The calculated losses at SINQ collimators show a 67$\%$ increase of the primary beam losses (from  4.3$ \pm 1.1\%$ at 0.5~mA to 7.2$ \pm 1.6\%$ at 1.86~mA). Such losses are consistent with the relative increase detected from surface temperature measurements as illustrated in Fig.~\ref{fig:KHE2_surface}.  
In order to verify the validity of the reported transmission drop with the current, we measured the inlet/outlet temperature of the heavy water cooling SINQ for two different beam currents of 0.51~mA and 1.93~mA. This yielded a power deposition at SINQ of 137.5~kW and 474.4~kW respectively. The normalized ratio of the power deposition with the current confirms the SINQ transmission drop from 0.5~mA to 1.9~mA.
Although remediating such losses is possible by focussing the beam onto the target more, as was confirmed by experiments, the target  should not become overheated and a balance between focus and beam losses needs to be found emperically.

\section{CONCLUSION}
Benchmarking the BDSIM/Geant4 simulations with the temperature measurements, beam profile measurements as well the beam current measurements allowed to understand and validate the accuracy of the Monte Carlo simulation tools. In particular, the intensity-dependent effects lead to continuously changing beam conditions at the target, and therefore to intensity-dependent power deposition and beam transmission. This justifies the importance of knowing the initial conditions of the beam impinging at the target. In addition, the newly established approach, based on measuring the variation of the surface temperature with the current, proved an effective tool to probe and understand the distributed losses of the MW-class beam: for instance, it was found out that KHE2 collimator and all 3 SINQ collimators (denoted KHN) play the most prominent role to explain the beam transmission evolution with the current, both qualitatively and quantitatively, as revealed by means of the relative excess power analysis. As a result, most losses in the beamline are controlled beam losses that explained the beam transmission drop with the current at different stages (MHC6b and SINQ). 
The good agreement between the water temperature and the surface temperature measurements confirm the validity of both approaches to monitor the average losses deposited by the high power beam at different locations. Therefore, installing temperature probes at the entrance/exit surface of the high power beamline elements can prove a cost-effective way to determine the localized losses and prevent long term damage to its key components. This is particularly relevant for particle accelerators aiming to deliver MW-class beams for Accelerator Driven Systems applications \cite{rubbia1995conceptual, bowman1998accelerator, abderrahim2012myrrha} or neutrino physics experiments \cite{winklehner2018high} and for careful monitoring of the beam transmission to the experiments. \\
Last but not least, owing to the successful benchmarking campaign with the existing SINQ beamline, we are confident that we can model and predict the beam losses for the HIPA upgrade project HIMB reliably as well. 

\begin{acknowledgments}
We acknowledge useful discussions with C. Baumgarten, H. Zhang, P.-A. Duperrex, R. D\"{o}lling, R. Sobbia and J. Kohlbrecher. Special thanks to L. Nevay and S.T. Boogert for helping with BDSIM and \textsc{Pyg4ometry}. We also thank J. Welte and C. Kramer for providing the information regarding the cooling circuit.
\end{acknowledgments}

\appendix

\section{New collimators for 3 mA operation} \label{app:A}

\begin{table}[htb]
\centering
\caption{Simulated power deposition in kW assuming a 3 mA proton beam current \label{Tab:power_depos_3mA}}
\def\arraystretch{1.2}
\begin{tabular}{lll}
\hline \hline
Beamline element&Present design&New design \\
\hline
KHE2&164.3&104.3\\

KHE3&32.2&40.0\\

$1^{st}$ quadrupole doublet&11.4&10.0\\

AHL magnet&11.4&15.3\\

$2^{nd}$ quadrupole doublet&$\sim 0$&$\sim 0$\\
\hline \hline
\end{tabular}
\end{table}
The simulated beam transmission $R_{\text{MHC6b}}$ at 3 mA shall drop to 68.6 $\pm$ 0.4\% with the existing collimator system. The newly designed one shows a larger transmission reaching 70.7 $\pm$ 0.4\%, i.e., nearly 2\% larger than the present system. Such an increase in transmission is due to the reduced beam losses with the new collimators design as is shown in Table~\ref{Tab:power_depos_3mA}. However, care must be paid to the increasing losses in the last part of the beamline. This is under investigation.

\nocite{*}

\bibliography{main}% Produces the bibliography via BibTeX.

\begin{thebibliography}{33}
\expandafter\ifx\csname natexlab\endcsname\relax\def\natexlab#1{#1}\fi
\expandafter\ifx\csname bibnamefont\endcsname\relax
  \def\bibnamefont#1{#1}\fi
\expandafter\ifx\csname bibfnamefont\endcsname\relax
  \def\bibfnamefont#1{#1}\fi
\expandafter\ifx\csname citenamefont\endcsname\relax
  \def\citenamefont#1{#1}\fi
\expandafter\ifx\csname url\endcsname\relax
  \def\url#1{\texttt{#1}}\fi
\expandafter\ifx\csname urlprefix\endcsname\relax\def\urlprefix{URL }\fi
\providecommand{\bibinfo}[2]{#2}
\providecommand{\eprint}[2][]{\url{#2}}

\bibitem[{\citenamefont{Grillenberger et~al.}(2021)\citenamefont{Grillenberger,
  Baumgarten, and Seidel}}]{grillenberger2021high}
\bibinfo{author}{\bibfnamefont{J.}~\bibnamefont{Grillenberger}},
  \bibinfo{author}{\bibfnamefont{C.}~\bibnamefont{Baumgarten}},
  \bibnamefont{and} \bibinfo{author}{\bibfnamefont{M.}~\bibnamefont{Seidel}},
  \bibinfo{journal}{SciPost Physics Proceedings} p. \bibinfo{pages}{002}
  (\bibinfo{year}{2021}).

\bibitem[{\citenamefont{Kiselev et~al.}(2021)\citenamefont{Kiselev, Duperrex,
  Jollet, Joray, Laube, Reggiani, Sobbia, and Talanov}}]{kiselev2021meson}
\bibinfo{author}{\bibfnamefont{D.}~\bibnamefont{Kiselev}},
  \bibinfo{author}{\bibfnamefont{P.-A.} \bibnamefont{Duperrex}},
  \bibinfo{author}{\bibfnamefont{S.}~\bibnamefont{Jollet}},
  \bibinfo{author}{\bibfnamefont{S.}~\bibnamefont{Joray}},
  \bibinfo{author}{\bibfnamefont{D.}~\bibnamefont{Laube}},
  \bibinfo{author}{\bibfnamefont{D.}~\bibnamefont{Reggiani}},
  \bibinfo{author}{\bibfnamefont{R.}~\bibnamefont{Sobbia}}, \bibnamefont{and}
  \bibinfo{author}{\bibfnamefont{V.}~\bibnamefont{Talanov}},
  \bibinfo{journal}{SciPost Physics Proceedings} p. \bibinfo{pages}{003}
  (\bibinfo{year}{2021}).

\bibitem[{\citenamefont{Eichler et~al.}(2022)}]{CDR_HIMB}
\bibinfo{author}{\bibfnamefont{R.}~\bibnamefont{Eichler}} \bibnamefont{et~al.},
  \bibinfo{journal}{Paul Scherrer Institut, Villigen PSI, Switzerland, Report
  No.: 22-01}  (\bibinfo{year}{2022}),
  \urlprefix\url{https://www.dora.lib4ri.ch/psi/islandora/object/psi:41209}.

\bibitem[{\citenamefont{PSI}(2022)}]{TATTOOS}
\bibinfo{author}{\bibnamefont{PSI}}, \emph{\bibinfo{title}{Tattoos
  description}} (\bibinfo{year}{2022}),
  \urlprefix\url{https://www.psi.ch/en/impact/tattoos}.

\bibitem[{\citenamefont{Aiba et~al.}(2021)\citenamefont{Aiba, Amato, Antognini,
  Ban, Berger, Caminada, Chislett, Crivelli, Crivellin, Maso
  et~al.}}]{aiba2021science}
\bibinfo{author}{\bibfnamefont{M.}~\bibnamefont{Aiba}},
  \bibinfo{author}{\bibfnamefont{A.}~\bibnamefont{Amato}},
  \bibinfo{author}{\bibfnamefont{A.}~\bibnamefont{Antognini}},
  \bibinfo{author}{\bibfnamefont{S.}~\bibnamefont{Ban}},
  \bibinfo{author}{\bibfnamefont{N.}~\bibnamefont{Berger}},
  \bibinfo{author}{\bibfnamefont{L.}~\bibnamefont{Caminada}},
  \bibinfo{author}{\bibfnamefont{R.}~\bibnamefont{Chislett}},
  \bibinfo{author}{\bibfnamefont{P.}~\bibnamefont{Crivelli}},
  \bibinfo{author}{\bibfnamefont{A.}~\bibnamefont{Crivellin}},
  \bibinfo{author}{\bibfnamefont{G.~D.} \bibnamefont{Maso}},
  \bibnamefont{et~al.}, \bibinfo{journal}{arXiv preprint arXiv:2111.05788}
  (\bibinfo{year}{2021}).

\bibitem[{\citenamefont{Berg et~al.}(2016)\citenamefont{Berg, Desorgher, Fuchs,
  Hajdas, Hodge, Kettle, Knecht, L{\"u}scher, Papa, Rutar
  et~al.}}]{berg2016target}
\bibinfo{author}{\bibfnamefont{F.}~\bibnamefont{Berg}},
  \bibinfo{author}{\bibfnamefont{L.}~\bibnamefont{Desorgher}},
  \bibinfo{author}{\bibfnamefont{A.}~\bibnamefont{Fuchs}},
  \bibinfo{author}{\bibfnamefont{W.}~\bibnamefont{Hajdas}},
  \bibinfo{author}{\bibfnamefont{Z.}~\bibnamefont{Hodge}},
  \bibinfo{author}{\bibfnamefont{P.-R.} \bibnamefont{Kettle}},
  \bibinfo{author}{\bibfnamefont{A.}~\bibnamefont{Knecht}},
  \bibinfo{author}{\bibfnamefont{R.}~\bibnamefont{L{\"u}scher}},
  \bibinfo{author}{\bibfnamefont{A.}~\bibnamefont{Papa}},
  \bibinfo{author}{\bibfnamefont{G.}~\bibnamefont{Rutar}},
  \bibnamefont{et~al.}, \bibinfo{journal}{Physical Review Accelerators and
  Beams} \textbf{\bibinfo{volume}{19}}, \bibinfo{pages}{024701}
  (\bibinfo{year}{2016}).

\bibitem[{\citenamefont{Nevay et~al.}(2020)\citenamefont{Nevay, Boogert,
  Snuverink, Abramov, Deacon, Garcia-Morales, Lefebvre, Gibson, Kwee-Hinzmann,
  Shields et~al.}}]{nevay2020bdsim}
\bibinfo{author}{\bibfnamefont{L.}~\bibnamefont{Nevay}},
  \bibinfo{author}{\bibfnamefont{S.}~\bibnamefont{Boogert}},
  \bibinfo{author}{\bibfnamefont{J.}~\bibnamefont{Snuverink}},
  \bibinfo{author}{\bibfnamefont{A.}~\bibnamefont{Abramov}},
  \bibinfo{author}{\bibfnamefont{L.}~\bibnamefont{Deacon}},
  \bibinfo{author}{\bibfnamefont{H.}~\bibnamefont{Garcia-Morales}},
  \bibinfo{author}{\bibfnamefont{H.}~\bibnamefont{Lefebvre}},
  \bibinfo{author}{\bibfnamefont{S.}~\bibnamefont{Gibson}},
  \bibinfo{author}{\bibfnamefont{R.}~\bibnamefont{Kwee-Hinzmann}},
  \bibinfo{author}{\bibfnamefont{W.}~\bibnamefont{Shields}},
  \bibnamefont{et~al.}, \bibinfo{journal}{Computer Physics Communications}
  \textbf{\bibinfo{volume}{252}}, \bibinfo{pages}{107200}
  (\bibinfo{year}{2020}).

\bibitem[{\citenamefont{Agostinelli et~al.}(2003)\citenamefont{Agostinelli,
  Allison, Amako, Apostolakis, Araujo, Arce, Asai, Axen, Banerjee, Barrand
  et~al.}}]{agostinelli2003geant4}
\bibinfo{author}{\bibfnamefont{S.}~\bibnamefont{Agostinelli}},
  \bibinfo{author}{\bibfnamefont{J.}~\bibnamefont{Allison}},
  \bibinfo{author}{\bibfnamefont{K.~a.} \bibnamefont{Amako}},
  \bibinfo{author}{\bibfnamefont{J.}~\bibnamefont{Apostolakis}},
  \bibinfo{author}{\bibfnamefont{H.}~\bibnamefont{Araujo}},
  \bibinfo{author}{\bibfnamefont{P.}~\bibnamefont{Arce}},
  \bibinfo{author}{\bibfnamefont{M.}~\bibnamefont{Asai}},
  \bibinfo{author}{\bibfnamefont{D.}~\bibnamefont{Axen}},
  \bibinfo{author}{\bibfnamefont{S.}~\bibnamefont{Banerjee}},
  \bibinfo{author}{\bibfnamefont{G.}~\bibnamefont{Barrand}},
  \bibnamefont{et~al.}, \bibinfo{journal}{Nuclear instruments and methods in
  physics research section A: Accelerators, Spectrometers, Detectors and
  Associated Equipment} \textbf{\bibinfo{volume}{506}}, \bibinfo{pages}{250}
  (\bibinfo{year}{2003}).

\bibitem[{\citenamefont{Lo~Meo et~al.}(2015)\citenamefont{Lo~Meo,
  Cort{\'e}s-Giraldo, Massimi, Lerendegui-Marco, Barbagallo, Colonna, Guerrero,
  Mancusi, Mingrone, Quesada et~al.}}]{lo2015geant4}
\bibinfo{author}{\bibfnamefont{S.}~\bibnamefont{Lo~Meo}},
  \bibinfo{author}{\bibfnamefont{M.}~\bibnamefont{Cort{\'e}s-Giraldo}},
  \bibinfo{author}{\bibfnamefont{C.}~\bibnamefont{Massimi}},
  \bibinfo{author}{\bibfnamefont{J.}~\bibnamefont{Lerendegui-Marco}},
  \bibinfo{author}{\bibfnamefont{M.}~\bibnamefont{Barbagallo}},
  \bibinfo{author}{\bibfnamefont{N.}~\bibnamefont{Colonna}},
  \bibinfo{author}{\bibfnamefont{C.}~\bibnamefont{Guerrero}},
  \bibinfo{author}{\bibfnamefont{D.}~\bibnamefont{Mancusi}},
  \bibinfo{author}{\bibfnamefont{F.}~\bibnamefont{Mingrone}},
  \bibinfo{author}{\bibfnamefont{J.}~\bibnamefont{Quesada}},
  \bibnamefont{et~al.}, \bibinfo{journal}{The European Physical Journal A}
  \textbf{\bibinfo{volume}{51}}, \bibinfo{pages}{1} (\bibinfo{year}{2015}).

\bibitem[{\citenamefont{Allison et~al.}(2016)}]{ALLISON2016186}
\bibinfo{author}{\bibfnamefont{J.}~\bibnamefont{Allison}} \bibnamefont{et~al.},
  \bibinfo{journal}{Nuclear Instruments and Methods in Physics Research Section
  A: Accelerators, Spectrometers, Detectors and Associated Equipment}
  \textbf{\bibinfo{volume}{835}}, \bibinfo{pages}{186} (\bibinfo{year}{2016}),
  ISSN \bibinfo{issn}{0168-9002},
  \urlprefix\url{https://www.sciencedirect.com/science/article/pii/S0168900216306957}.

\bibitem[{\citenamefont{Lechner et~al.}(2019)\citenamefont{Lechner, Auchmann,
  Baer, Castro, Bruce, Cerutti, Esposito, Ferrari, Jowett, Mereghetti
  et~al.}}]{lechner2019validation}
\bibinfo{author}{\bibfnamefont{A.}~\bibnamefont{Lechner}},
  \bibinfo{author}{\bibfnamefont{B.}~\bibnamefont{Auchmann}},
  \bibinfo{author}{\bibfnamefont{T.}~\bibnamefont{Baer}},
  \bibinfo{author}{\bibfnamefont{C.~B.} \bibnamefont{Castro}},
  \bibinfo{author}{\bibfnamefont{R.}~\bibnamefont{Bruce}},
  \bibinfo{author}{\bibfnamefont{F.}~\bibnamefont{Cerutti}},
  \bibinfo{author}{\bibfnamefont{L.}~\bibnamefont{Esposito}},
  \bibinfo{author}{\bibfnamefont{A.}~\bibnamefont{Ferrari}},
  \bibinfo{author}{\bibfnamefont{J.}~\bibnamefont{Jowett}},
  \bibinfo{author}{\bibfnamefont{A.}~\bibnamefont{Mereghetti}},
  \bibnamefont{et~al.}, \bibinfo{journal}{Physical Review Accelerators and
  Beams} \textbf{\bibinfo{volume}{22}}, \bibinfo{pages}{071003}
  (\bibinfo{year}{2019}).

\bibitem[{\citenamefont{Grote and Schmidt}(2003)}]{grote2003mad}
\bibinfo{author}{\bibfnamefont{H.}~\bibnamefont{Grote}} \bibnamefont{and}
  \bibinfo{author}{\bibfnamefont{F.}~\bibnamefont{Schmidt}}, in
  \emph{\bibinfo{booktitle}{Proceedings of the 2003 Particle Accelerator
  Conference}} (\bibinfo{organization}{IEEE}, \bibinfo{year}{2003}),
  vol.~\bibinfo{volume}{5}, pp. \bibinfo{pages}{3497--3499}.

\bibitem[{GDM(2022)}]{GDML}
\emph{\bibinfo{title}{Gdml description website}} (\bibinfo{year}{2022}),
  \urlprefix\url{https://gdml.web.cern.ch/GDML/}.

\bibitem[{\citenamefont{Walker et~al.}(2022)\citenamefont{Walker, Abramov,
  Nevay, Shields, and Boogert}}]{WALKER2022108228}
\bibinfo{author}{\bibfnamefont{S.}~\bibnamefont{Walker}},
  \bibinfo{author}{\bibfnamefont{A.}~\bibnamefont{Abramov}},
  \bibinfo{author}{\bibfnamefont{L.}~\bibnamefont{Nevay}},
  \bibinfo{author}{\bibfnamefont{W.}~\bibnamefont{Shields}}, \bibnamefont{and}
  \bibinfo{author}{\bibfnamefont{S.}~\bibnamefont{Boogert}},
  \bibinfo{journal}{Computer Physics Communications}
  \textbf{\bibinfo{volume}{272}}, \bibinfo{pages}{108228}
  (\bibinfo{year}{2022}), ISSN \bibinfo{issn}{0010-4655},
  \urlprefix\url{https://www.sciencedirect.com/science/article/pii/S0010465521003404}.

\bibitem[{\citenamefont{Brown et~al.}(1980)\citenamefont{Brown, Rothacker,
  Carey, and Iselin}}]{brown1980transport}
\bibinfo{author}{\bibfnamefont{K.~L.} \bibnamefont{Brown}},
  \bibinfo{author}{\bibfnamefont{F.}~\bibnamefont{Rothacker}},
  \bibinfo{author}{\bibfnamefont{D.~C.} \bibnamefont{Carey}}, \bibnamefont{and}
  \bibinfo{author}{\bibfnamefont{C.}~\bibnamefont{Iselin}},
  \bibinfo{type}{Tech. Rep.}, \bibinfo{institution}{European Organization for
  Nuclear Research} (\bibinfo{year}{1980}).

\bibitem[{\citenamefont{Agapov}(2006)}]{agapov2006gmad}
\bibinfo{author}{\bibfnamefont{I.}~\bibnamefont{Agapov}}, \bibinfo{type}{Tech.
  Rep.} (\bibinfo{year}{2006}).

\bibitem[{\citenamefont{Rizzoglio et~al.}(2017)\citenamefont{Rizzoglio,
  Adelmann, Baumgarten, Frey, Gerbershagen, Meer, and
  Schippers}}]{rizzoglio2017evolution}
\bibinfo{author}{\bibfnamefont{V.}~\bibnamefont{Rizzoglio}},
  \bibinfo{author}{\bibfnamefont{A.}~\bibnamefont{Adelmann}},
  \bibinfo{author}{\bibfnamefont{C.}~\bibnamefont{Baumgarten}},
  \bibinfo{author}{\bibfnamefont{M.}~\bibnamefont{Frey}},
  \bibinfo{author}{\bibfnamefont{A.}~\bibnamefont{Gerbershagen}},
  \bibinfo{author}{\bibfnamefont{D.}~\bibnamefont{Meer}}, \bibnamefont{and}
  \bibinfo{author}{\bibfnamefont{J.}~\bibnamefont{Schippers}},
  \bibinfo{journal}{Physical Review Accelerators and Beams}
  \textbf{\bibinfo{volume}{20}}, \bibinfo{pages}{124702}
  (\bibinfo{year}{2017}).

\bibitem[{\citenamefont{Stammbach et~al.}(2001)\citenamefont{Stammbach, Adam,
  Blumer, George, Mezger, Schmelzbach, and Sigg}}]{stammbach2001psi}
\bibinfo{author}{\bibfnamefont{T.}~\bibnamefont{Stammbach}},
  \bibinfo{author}{\bibfnamefont{S.}~\bibnamefont{Adam}},
  \bibinfo{author}{\bibfnamefont{T.}~\bibnamefont{Blumer}},
  \bibinfo{author}{\bibfnamefont{D.}~\bibnamefont{George}},
  \bibinfo{author}{\bibfnamefont{A.}~\bibnamefont{Mezger}},
  \bibinfo{author}{\bibfnamefont{P.}~\bibnamefont{Schmelzbach}},
  \bibnamefont{and} \bibinfo{author}{\bibfnamefont{P.}~\bibnamefont{Sigg}}, in
  \emph{\bibinfo{booktitle}{AIP Conference Proceedings}}
  (\bibinfo{organization}{American Institute of Physics},
  \bibinfo{year}{2001}), vol. \bibinfo{volume}{600}, pp.
  \bibinfo{pages}{423--427}.

\bibitem[{\citenamefont{Baartman}(2013)}]{baartman2013space}
\bibinfo{author}{\bibfnamefont{R.}~\bibnamefont{Baartman}}, \bibinfo{journal}{a
  (a+ b)} \textbf{\bibinfo{volume}{10}}, \bibinfo{pages}{2}
  (\bibinfo{year}{2013}).

\bibitem[{\citenamefont{Kolano et~al.}(2018)\citenamefont{Kolano, Adelmann,
  Barlow, and Baumgarten}}]{kolano2018intensity}
\bibinfo{author}{\bibfnamefont{A.}~\bibnamefont{Kolano}},
  \bibinfo{author}{\bibfnamefont{A.}~\bibnamefont{Adelmann}},
  \bibinfo{author}{\bibfnamefont{R.}~\bibnamefont{Barlow}}, \bibnamefont{and}
  \bibinfo{author}{\bibfnamefont{C.}~\bibnamefont{Baumgarten}},
  \bibinfo{journal}{Nuclear Instruments and Methods in Physics Research Section
  A: Accelerators, Spectrometers, Detectors and Associated Equipment}
  \textbf{\bibinfo{volume}{885}}, \bibinfo{pages}{54} (\bibinfo{year}{2018}).

\bibitem[{\citenamefont{Stetson et~al.}(1992)\citenamefont{Stetson, Adam,
  Humbel, Joho, and Stammbach}}]{bib:stetsoncommissioning}
\bibinfo{author}{\bibfnamefont{J.}~\bibnamefont{Stetson}},
  \bibinfo{author}{\bibfnamefont{S.}~\bibnamefont{Adam}},
  \bibinfo{author}{\bibfnamefont{M.}~\bibnamefont{Humbel}},
  \bibinfo{author}{\bibfnamefont{W.}~\bibnamefont{Joho}}, \bibnamefont{and}
  \bibinfo{author}{\bibfnamefont{T.}~\bibnamefont{Stammbach}}
  (\bibinfo{year}{1992}).

\bibitem[{\citenamefont{Rossi and Greisen}(1941)}]{rossi1941cosmic}
\bibinfo{author}{\bibfnamefont{B.}~\bibnamefont{Rossi}} \bibnamefont{and}
  \bibinfo{author}{\bibfnamefont{K.}~\bibnamefont{Greisen}},
  \bibinfo{journal}{Reviews of Modern Physics} \textbf{\bibinfo{volume}{13}},
  \bibinfo{pages}{240} (\bibinfo{year}{1941}).

\bibitem[{\citenamefont{Highland}(1975)}]{highland1975some}
\bibinfo{author}{\bibfnamefont{V.~L.} \bibnamefont{Highland}},
  \bibinfo{journal}{Nuclear Instruments and Methods}
  \textbf{\bibinfo{volume}{129}}, \bibinfo{pages}{497} (\bibinfo{year}{1975}).

\bibitem[{\citenamefont{Banerjee}(2005)}]{bib:banerjee2005evaluation}
\bibinfo{author}{\bibfnamefont{B.}~\bibnamefont{Banerjee}},
  \bibinfo{journal}{arXiv preprint cond-mat/0512466}  (\bibinfo{year}{2005}).

\bibitem[{\citenamefont{Seidel and Schmelzbach}(2007)}]{seidel2007upgrade}
\bibinfo{author}{\bibfnamefont{M.}~\bibnamefont{Seidel}} \bibnamefont{and}
  \bibinfo{author}{\bibfnamefont{P.}~\bibnamefont{Schmelzbach}},
  \bibinfo{journal}{Proc. Cycl. and their Appl}  (\bibinfo{year}{2007}).

\bibitem[{\citenamefont{Lee et~al.}(2010)\citenamefont{Lee, Reggiani, Gandel,
  Kiselev, Baumann, and Seidel}}]{lee2010new}
\bibinfo{author}{\bibfnamefont{Y.}~\bibnamefont{Lee}},
  \bibinfo{author}{\bibfnamefont{D.}~\bibnamefont{Reggiani}},
  \bibinfo{author}{\bibfnamefont{M.}~\bibnamefont{Gandel}},
  \bibinfo{author}{\bibfnamefont{D.}~\bibnamefont{Kiselev}},
  \bibinfo{author}{\bibfnamefont{P.}~\bibnamefont{Baumann}}, \bibnamefont{and}
  \bibinfo{author}{\bibfnamefont{M.}~\bibnamefont{Seidel}},
  \bibinfo{journal}{HB2010, Morschach, Switzerland}  (\bibinfo{year}{2010}).

\bibitem[{\citenamefont{Duperrex et~al.}(2010)\citenamefont{Duperrex, Gandel,
  Kiselev, Lee, M{\"u}ller et~al.}}]{duperrex2010current}
\bibinfo{author}{\bibfnamefont{P.-A.} \bibnamefont{Duperrex}},
  \bibinfo{author}{\bibfnamefont{M.}~\bibnamefont{Gandel}},
  \bibinfo{author}{\bibfnamefont{D.}~\bibnamefont{Kiselev}},
  \bibinfo{author}{\bibfnamefont{Y.}~\bibnamefont{Lee}},
  \bibinfo{author}{\bibfnamefont{U.}~\bibnamefont{M{\"u}ller}},
  \bibnamefont{et~al.}, \bibinfo{journal}{Proc. HB} pp.
  \bibinfo{pages}{443--447} (\bibinfo{year}{2010}).

\bibitem[{\citenamefont{Reggiani et~al.}(2018)\citenamefont{Reggiani,
  D{\"o}lling, Duperrex, Kiselev, Welte, and
  Wohlmuther}}]{reggiani2018improving}
\bibinfo{author}{\bibfnamefont{D.}~\bibnamefont{Reggiani}},
  \bibinfo{author}{\bibfnamefont{R.}~\bibnamefont{D{\"o}lling}},
  \bibinfo{author}{\bibfnamefont{P.-A.} \bibnamefont{Duperrex}},
  \bibinfo{author}{\bibfnamefont{D.}~\bibnamefont{Kiselev}},
  \bibinfo{author}{\bibfnamefont{J.}~\bibnamefont{Welte}}, \bibnamefont{and}
  \bibinfo{author}{\bibfnamefont{M.}~\bibnamefont{Wohlmuther}},
  \bibinfo{journal}{Proc. IPAC'18} pp. \bibinfo{pages}{2337--2340}
  (\bibinfo{year}{2018}).

\bibitem[{\citenamefont{Duperrex}()}]{Priv_Comm_PA}
\bibinfo{author}{\bibfnamefont{P.-A.} \bibnamefont{Duperrex}},
  \bibinfo{howpublished}{{Private Communication}}.

\bibitem[{\citenamefont{Rubbia et~al.}(1995)\citenamefont{Rubbia, Roche, Rubio,
  Carminati, Kadi, Mandrillon, Revol, Buono, Klapisch, Fi{\'e}tier
  et~al.}}]{rubbia1995conceptual}
\bibinfo{author}{\bibfnamefont{C.}~\bibnamefont{Rubbia}},
  \bibinfo{author}{\bibfnamefont{C.}~\bibnamefont{Roche}},
  \bibinfo{author}{\bibfnamefont{J.~A.} \bibnamefont{Rubio}},
  \bibinfo{author}{\bibfnamefont{F.}~\bibnamefont{Carminati}},
  \bibinfo{author}{\bibfnamefont{Y.}~\bibnamefont{Kadi}},
  \bibinfo{author}{\bibfnamefont{P.}~\bibnamefont{Mandrillon}},
  \bibinfo{author}{\bibfnamefont{J.~P.~C.} \bibnamefont{Revol}},
  \bibinfo{author}{\bibfnamefont{S.}~\bibnamefont{Buono}},
  \bibinfo{author}{\bibfnamefont{R.}~\bibnamefont{Klapisch}},
  \bibinfo{author}{\bibfnamefont{N.}~\bibnamefont{Fi{\'e}tier}},
  \bibnamefont{et~al.}, \bibinfo{type}{Tech. Rep.} (\bibinfo{year}{1995}).

\bibitem[{\citenamefont{Bowman}(1998)}]{bowman1998accelerator}
\bibinfo{author}{\bibfnamefont{C.~D.} \bibnamefont{Bowman}},
  \bibinfo{journal}{Annual Review of Nuclear and Particle Science}
  \textbf{\bibinfo{volume}{48}}, \bibinfo{pages}{505} (\bibinfo{year}{1998}).

\bibitem[{\citenamefont{Abderrahim et~al.}(2012)\citenamefont{Abderrahim,
  Baeten, De~Bruyn, and Fernandez}}]{abderrahim2012myrrha}
\bibinfo{author}{\bibfnamefont{H.~A.} \bibnamefont{Abderrahim}},
  \bibinfo{author}{\bibfnamefont{P.}~\bibnamefont{Baeten}},
  \bibinfo{author}{\bibfnamefont{D.}~\bibnamefont{De~Bruyn}}, \bibnamefont{and}
  \bibinfo{author}{\bibfnamefont{R.}~\bibnamefont{Fernandez}},
  \bibinfo{journal}{Energy conversion and management}
  \textbf{\bibinfo{volume}{63}}, \bibinfo{pages}{4} (\bibinfo{year}{2012}).

\bibitem[{\citenamefont{Winklehner et~al.}(2018)\citenamefont{Winklehner,
  Bahng, Calabretta, Calanna, Chakrabarti, Conrad, D’Agostino, Dechoudhury,
  Naik, Waites et~al.}}]{winklehner2018high}
\bibinfo{author}{\bibfnamefont{D.}~\bibnamefont{Winklehner}},
  \bibinfo{author}{\bibfnamefont{J.}~\bibnamefont{Bahng}},
  \bibinfo{author}{\bibfnamefont{L.}~\bibnamefont{Calabretta}},
  \bibinfo{author}{\bibfnamefont{A.}~\bibnamefont{Calanna}},
  \bibinfo{author}{\bibfnamefont{A.}~\bibnamefont{Chakrabarti}},
  \bibinfo{author}{\bibfnamefont{J.}~\bibnamefont{Conrad}},
  \bibinfo{author}{\bibfnamefont{G.}~\bibnamefont{D’Agostino}},
  \bibinfo{author}{\bibfnamefont{S.}~\bibnamefont{Dechoudhury}},
  \bibinfo{author}{\bibfnamefont{V.}~\bibnamefont{Naik}},
  \bibinfo{author}{\bibfnamefont{L.}~\bibnamefont{Waites}},
  \bibnamefont{et~al.}, \bibinfo{journal}{Nuclear Instruments and Methods in
  Physics Research Section A: Accelerators, Spectrometers, Detectors and
  Associated Equipment} \textbf{\bibinfo{volume}{907}}, \bibinfo{pages}{231}
  (\bibinfo{year}{2018}).

\end{thebibliography}

\end{document}